\documentclass[useAMS,usenatbib]{mn2e}
\pdfoutput=1
\usepackage{graphics}
\usepackage{graphicx}
\usepackage{amsxtra}
\usepackage{array,longtable}
\usepackage{journals}
\usepackage{url}

\usepackage{latexsym}
\usepackage{amssymb}

\usepackage{ulem}
\usepackage{color}
\newcommand{\ser}{S\'ersic}

\title
[Stellar disc flattening]
{Does the stellar disc flattening depend on the galaxy type?
}
\author[A.V.~Mosenkov, N.Ya.~Sotnikova, V.P.~Reshetnikov, D.V.~Bizyaev, and S.J.~Kautsch]
{A.V.~Mosenkov$^{1,2,3}$\thanks{E-mail: mosenkovAV@gmail.com}, N.Ya.~Sotnikova$^{3,4}$, V.P.~Reshetnikov$^{3,4}$, D.V.~Bizyaev$^{5,6}$, \newauthor and S.J.~Kautsch$^{7}$
\\
$^1$Sterrenkundig Observatorium, Universiteit Gent, Krijgslaan 281 S9,
B-9000 Gent, Belgium\\
$^2$Central Astronomical Observatory of RAS, Russia\\
$^3$St.Petersburg State University, Universitetskij pr. 28, 198504
St.Petersburg, Stary Peterhof, Russia \\
$^4$Isaac Newton Institute of Chile, St.Petersburg Branch\\
$^5$Apache Point Observatory and New Mexico State University, Sunspot, NM, 88349, USA\\
$^6$Sternberg Astronomical Institute, Moscow State University, Moscow, Russia\\
$^7$Nova Southeastern University, Fort Lauderdale, FL,
33314, USA
}
\begin{document}
%%%%%%%%%%%%%%%%%%%%%%%%%%%%%%%%%%%%%%%%%%%%%%%%%%%%%%%%%%%%%%%%%%%%%%

\date{Accepted 2015 ???. Received ???; in original form 2015 ???}

\pagerange{\pageref{firstpage}--\pageref{lastpage}} \pubyear{2015}

\maketitle

\label{firstpage}

%%%%%%%%%%%%%%%%%%%%%%%%%%%%%%%%%%%%%%%%%%%%%%%%%%%%%%%%%%%%%%%%%%%%%%
\begin{abstract}
%%%%%%%%%%%%%%%%%%%%%%%%%%%%%%%%%%%%%%%%%%%%%%%%%%%%%%%%%%%%%%%%%%%%%%

We analyze the dependence of the stellar disc flatness on the 
galaxy morphological type using 2D decomposition of galaxies 
from the reliable subsample of the Edge-on Galaxies in SDSS (EGIS) 
catalogue. Combining these data with the retrieved models of the 
edge-on galaxies from the Two Micron All Sky Survey (2MASS) and 
the Spitzer Survey of Stellar Structure in Galaxies (S$^4$G) 
catalogue, we make the following conclusions: 

(1) The disc relative thickness $z_0/h$ 
in the near- and mid-infrared passbands
correlates weakly 
with morphological type  and 
does not correlate with the bulge-to-total luminosity ratio 
$B/T$ in all studied bands. 

(2) Applying an 1D photometric profile analysis overestimates the disc 
thickness in galaxies with large bulges making an illusion of the 
relationship between the disc flattening and the ratio $B/T$. 

(3) In our sample the early-type disc galaxies (S0/a) have both 
flat and ``puffed'' discs. The early spirals and intermediate-type 
galaxies have a large scatter of the disc flatness, which can be 
caused by the presence of a bar: barred galaxies have thicker 
stellar discs, on average. On the other hand, the late-type spirals 
are mostly thin galaxies, whereas irregular galaxies have puffed 
stellar discs. 

\end{abstract}


%%%%%%%%%%%%%%%%%%%%%%%%%%%%%%%%%%%%%%%%%%%%%%%%%%%%%%%%%%%%%%%%%%%%%%
\section{Introduction}
%%%%%%%%%%%%%%%%%%%%%%%%%%%%%%%%%%%%%%%%%%%%%%%%%%%%%%%%%%%%%%%%%%%%%%
Morphological and photometric characteristics of galaxies vary along 
the Hubble sequence (see the review by \citealp{Roberts+1994}). 
Contribution of the bulge to the total luminosity of a galaxy 
decreases, galaxies tend to be bluer, spiral pattern gets less 
tightly wound, the fraction of gas increases while going from early- 
to late-type galaxies. 

\cite{Guthrie1992} analyzed the distribution of isophotal axial 
ratios in a large sample of UGC galaxies to study 
true axial ratios $b/a$ and found the dependence of the true 
axial ratio on the morphological type and that the late type galaxies 
have smaller $b/a$. \cite{Karachentsev+1997} measured the 
apparent axial ratios $b/a$ of galaxies from the Flat Galaxies 
Catalog \citep{Karachentsev+1993} and found that the flat galaxies 
exhibit the dependence of the flattening on the morphological type: 
galaxies of the late types are flatter. 

Although the apparent axial ratio is an indicator of the disc 
thickness, these two characteristics are not the same. 
\cite{deGrijs1998} performed the 1D photometric profile analysis 
for 45 edge-on galaxies and derived global galaxy parameters in 
three photometric bands (optical and near-infrared). He found a 
correlation between the ratio of the radial to vertical scale 
parameter and galaxy type: galaxies become systematically thinner 
when going from the type S0 to Sc. According to \cite{deGrijs1998}, 
the ratio of the radial scale length $h$ to the vertical scale height 
$z_0$ in his sample of edge-on galaxies varies from 1.5--2 in the 
early-type spirals to 3--8 in the Sc--Sd galaxies. Note, however, 
that the scatter in $z_0/h$ is large for any certain morphological 
type.

\cite{Kregel+2002} re-analyzed the $I$ band photometry of 
34 edge-on spiral galaxies from the sample by \cite{deGrijs1998}. 
They applied a 2D algorithm to derive the scale parameters of the 
stellar disc and, thereby, the disc flattening. They also used the 
revised Hubble types of the galaxies from the HyperLeda database,
which were significantly shifted around as compared with those used 
by \cite{deGrijs1998}, sometimes by more than one subtype. 
However, \cite{Kregel+2002} did not use the revised data to discuss 
the dependence of the ratio $z_0/h$ on the morphological type found 
earlier by \cite{deGrijs1998} on the base of the same sample. It was 
done by \cite{Hernandez+2006}, who analyzed the data by 
\cite{Kregel+2002} and confirmed a diminishing trend of the ratio 
$z_0/h$ while going from the early to late types: the late-type 
galaxies at the Sc end have thinner discs than the earlier types, 
on average. Careful examination of fig.~7 
in \cite{Hernandez+2006} shows that the trend is supported by only 
one point (a galaxy of Sa type). 
Therefore, we can conclude that the reported correlation is rather poor.

\cite{Ma+1997,Ma+1998,Ma+1999} explored some statistical 
correlations for spiral galaxies of intermediate inclinations. 
They measured the disc thickness by the method proposed by 
\cite{Peng1988}. The method for determining the thickness of a 
spiral galaxy is based on the solution of Poisson's equation for 
a logarithmic disturbance of density. \cite{Ma+1997,Ma+1998,Ma+1999} 
showed that the ratio of the vertical to radial scales in the 
stellar disc tends to decrease along the Hubble sequence. 
It is important to note that all mentioned results are based on 
indirect methods for determining the disc flatness, on a simple 
1D photometric profile analysis or on limited data sets. 

\begin{figure}
\includegraphics[width=8.0cm, angle=0, clip=]{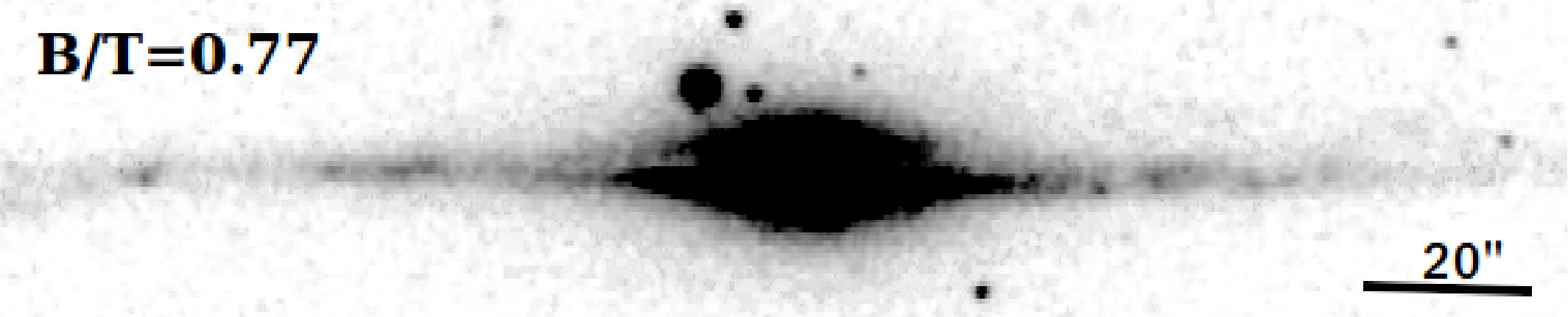}
\includegraphics[width=8.0cm, angle=0, clip=]{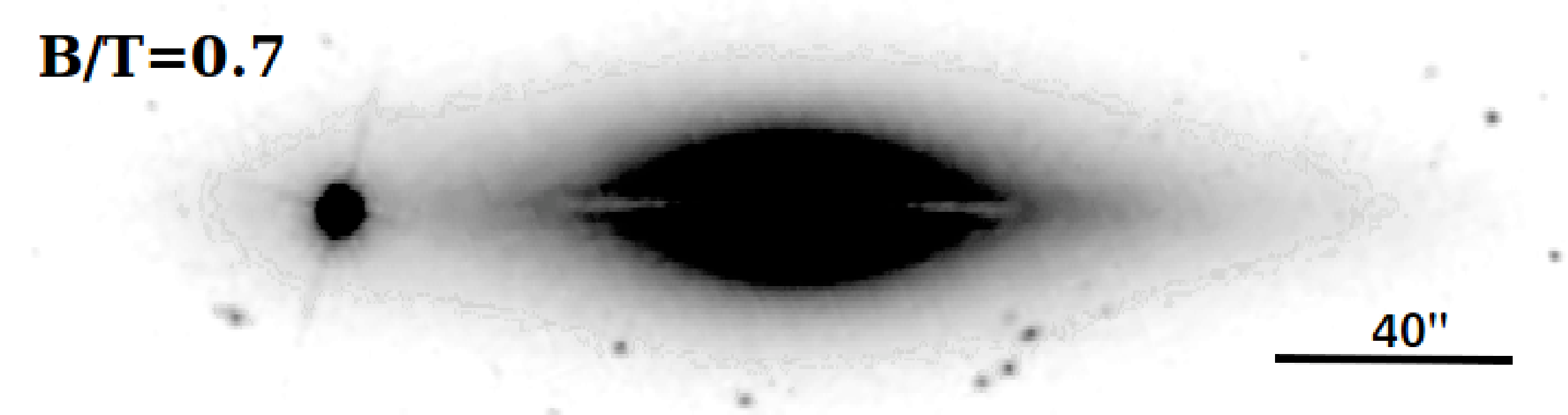}
\caption{Examples of galaxies from the EGIS catalogue ($i$ band): 
 IC~3608, Sb (Sab as given in the Catalogue of Detailed Visual 
 Morphological Classifications for 14,034 Galaxies in SDSS, 
 \citealp{Nair+2010}), $z_0/h=0.10$, bulge-to-total luminosity 
 ratio $B/T=0.77$ (top); and NGC~1032, S0/a 
 (Sa in the Uppsala General Catalogue of Galaxies, 
 \citealp{Nilson1973}), $z_0/h=0.30$, $B/T=0.70$ (bottom). 
 The morphological types are taken from the HyperLeda database. 
 The parameters are estimated via the 2D bulge/disc decomposition 
 using DECA package \citep{mos2014} with masking the central dust lanes 
 (see Section~\ref{SDSS} and also Fig.~\ref{ex_dec}).}
\label{galaxies_ex}
\end{figure}

\citet{Bizyaev+2014} (BKM14 hereafter) created the catalogue EGIS, 
a catalogue of edge-on galaxies in the Sloan Digital Sky Survey 
(SDSS). In their 3D photometric analysis of 5747 edge-on galaxies 
selected from SDSS Data Release 8 they did not find significant 
correlation between the disc flatness and the morphological type. 
Moreover, a visual inspection of galaxy images shows that early-type 
galaxies may have very flattened, as well as ``puffed" discs 
(see Fig.~\ref{galaxies_ex}). \cite{BizyaevKajsin2004} also noticed
the existence of very thin stellar disks in galaxies with large bulges.

To sort out this question, we analyze the dependence of the stellar 
disc flatness on the galaxy morphological type using robust 2D 
decomposition results obtained for a representative subsample of the EGIS 
catalogue. We also use the results of studying 
edge-on galaxies from the 2MASS archive (\citealp{Mosenkov+2010}, 
MSR10 hereafter) and from the Spitzer Survey of Stellar Structure 
in Galaxies (S$^4$G, \citealp{sheth+2010}).

This paper is organized as follows. 
In Section~\ref{Techniques}, we describe our methodology of the 2D
bulge/disc decomposition. We show that the approach used for 
decomposing edge-on galaxies brings up statistically reliable results. 
In Section~\ref{s_samples}, we introduce the samples of galaxies 
which is considered in this work. 
In Section~\ref{Results}, we present the results of our analysis, 
namely the correlations between the disc flatness and other 
quantities such as the morphological type, bulge-to-total 
luminosity ratio, galaxy colour, and galaxy luminosity. 
In Section~\ref{Discussion}, we discuss why the results of this 
article contradict the results of previous investigations. 
We also make some comments on the physical sense of the lack of 
the correlation between the disc flattening and the galaxy 
morphological type. 
In Section~\ref{Conclusions}, we summarize our main conclusions. 
The cosmological framework adopted throughout this paper is 
$H_0=70$~km\,s$^{-1}$\,Mpc$^{-1}$, $\Omega_m=0.3$, 
and $\Omega_\Lambda=0.7$. 
%%%%%%%%%%%%%%%%%%%%%%%%%%%%%%%%%%%%%%%%%%%%%%%%%%%%%%%%%%%%%%%%%%%%%%
%%%% Fig_1
%%%%%%%%%%%%%%%%%%%%%%%%%%%%%%%%%%%%%%%%%%%%%%%%%%%%%%%%%%%%%%%%%%%%%%
%%%%%%%%%%%%%%%%%%%%%%%%%%%%%%%%%%%%%%%%%%%%%%%%%%%%%%%%%%%%%%%%%%%%%%

%%%%%%%%%%%%%%%%%%%%%%%%%%%%%%%%%%%%%%%%%%%%%%%%%%%%%%%%%%%%%%%%%%%%%%
\section{The decomposition technique}
\label{Techniques}
%%%%%%%%%%%%%%%%%%%%%%%%%%%%%%%%%%%%%%%%%%%%%%%%%%%%%%%%%%%%%%%%%%%%%%
%In this work we use three approaches to decompose galaxy images 
%{\bf which are widely used in the literature to derive parameters 
%of galaxy components}. Here we briefly describe these techniques 
%focusing on edge-on galaxies and addressing the reader to the 
%respective articles where these techniques are given in details. 
We first describe the main decomposition technique employed in this 
study to analyze the structure of edge-on galaxies. There are three 
methods for decomposing such objects widely used in 
the structural analysis: the one-dimensional (1D, e.g. 
\citealp{deGrijs1998,biz2002} and BKM14), two-dimensional 
(2D, see \citealp{Simard+2002,Peng+2010,desouza+2004}), 
and three-dimensional method (3D, see, for example, 
\citealp{Xilouris+1999,bianchi07,Baes+2010} and BKM14). 
% incorporated into many public packages for photometric decomposition 
%(for example, {\sevensize GIM2D}, \citealp{Simard+2002}; {\sevensize GALFIT}, \citealp{Peng+2002,Peng+2010}; {\sevensize BUDDA}, \citealp{desouza+2004}).
%\citep[see][]{Mosenkov+2010,mos2014}. 

The 1D methods applied to edge-on galaxies considers photometric 
profiles 
along the major and minor axes independently. In contrast to that, 
in 2D fitting the whole information from the 2D image 
is used simultaneously to build a robust model for each galactic
component. There are several examples in the literature showing 
that the 2D method is much more reliable than the 1D method 
\citep[e.g.][]{dejong1996} in retrieving the accurate structural 
parameters. In the 3D method a dust extinction model is usually 
adopted, which makes this method more precise but rather 
time consuming. Since proper treatment of radiative transfer models 
takes significant computational resources \citep{bianchi07,Baes+2010,DeLooze+2012},
various simplifications can be applied (BKM14).
In this section, we briefly describe the adopted 2D 
technique in the comparison with the 1D and 3D methods in 
Section~\ref{diff_dec}.

%%%%%%%%%%%%%%%%%%%%%%%%%%%%%%%%%%%%%%%%%%%%%%%%%%%%%%%%%%%%%%%%%%%%%%
\subsection{The 2D decomposition}
\label{2D method description}
%%%%%%%%%%%%%%%%%%%%%%%%%%%%%%%%%%%%%%%%%%%%%%%%%%%%%%%%%%%%%%%%%%%%%%
The 2D photometric decomposition of galactic images is realized in a number of 
packages such as {\sevensize GIM2D} \citep{Simard+2002}, 
{\sevensize GALFIT} \citep{Peng+2002,Peng+2010}, 
{\sevensize BUDDA} \citep{desouza+2004}, and 
{\sevensize IMFIT} \citep{Erwin2014}. In this work 
we use three of them: {\sevensize BUDDA}, {\sevensize IMFIT}, 
and the most famous code {\sevensize GALFIT}. Although the 
minimization algorithms are different in these packages, we assume 
they bring similar results of the decomposition since they have been tested and 
compared by different authors 
\citep[e.g.][]{Haussler+2007,Haussler+2013,Kim+2014,Busch+2014}. 
Also, each of these codes has its advantages which will be noticed 
further.

To perform the bulge/disc decomposition, new Python wrapper 
{\sevensize DECA}\footnote{http://lacerta.astro.spbu.ru/?q=node/96} was made, as described in detail by 
\cite{mos2014}. 
The {\sevensize DECA} software package was designed to perform 
photometric analysis of the structure of regular galaxies, with 
the simple automatic decomposition onto bulge and disc using 
{\sevensize GALFIT} as the galactic model generator. It also takes advantages 
of several widely used packages such as {\sevensize IRAF}, 
{\sevensize SExtractor} and several Python libraries as, 
for example, {\sevensize PyFITS}. 
{\sevensize DECA} requires minimal human intervention as no any 
initial values of the bulge or disc parameters should be given. 
The algorithm is built in the way that it uses the input image 
of the studied galaxy (or, it may be a field with many different 
objects), the Point Spread Function (PSF) image, and the input 
text file with the main information about the object (equatorial 
coordinates RA and DEC, redshift, and colour) and the frame calibration 
(photometric zero-point, pixel scale, gain, etc). 
%The {\sevensize DECA} package combines different procedures to 
%perform special tasks executed in the following sequence:
%\begin{itemize}
%\item[$\cdot$] Removal or masking of ``contaminants'' (objects that 
%do not belong to the galaxy studied, e.g. other galaxies, stars, 
%globular clusters or CCD artifacts).
%\item[$\cdot$] Correction for the sky background. 
%\item[$\cdot$] Taking into account atmospheric and optical system 
%effects represented with the PSF.
%\item[$\cdot$] Accurate cropping of the galaxy image for subsequent 
%analysis. 
%\item[$\cdot$] Selection of initial approximations of component 
%parameters (initial guess) that are as close to reality as possible. 
%\end{itemize}
%All these steps are explained in detail in \cite{mos2014}. In Section~\ref{} we make some comments to each of them when we apply the {\sevensize DECA} package to a sample of edge-on galaxies.

A simple model of an edge-on disc galaxy represents a superposition 
of two photometric components: the bulge and the edge-on disc. 
The distribution of the surface brightness in the radial $r$ and vertical $z$ 
directions for the transparent `exponential' disc observed at the edge-on 
orientation is described by the following expression:
%%%%%%%%%%%%%%%%%%%%%%%%%%%%%%%%%%%%%%%%%%%%%%%%%%%%%%%%%%%%%%%%%%%%%%
\begin{equation}
I(r,z) = 
  I(0,0)
  \displaystyle \frac{r}{h} \, K_1\left(\frac{r}{h}\right) 
  \mathrm{sech}^2(z/z_0) , \,
\label{form_DSB}
\end{equation}
%%%%%%%%%%%%%%%%%%%%%%%%%%%%%%%%%%%%%%%%%%%%%%%%%%%%%%%%%%%%%%%%%%%%%%
where $I(0,0)$ is the disc central intensity, $h$ is the radial 
scale length, $z_0$ is the `isothermal' scale height of the disc 
\citep{Spitzer1942,vanderKruit1981a,vanderKruit1981b,vanderKruit1982a,vanderKruit1982b},
and $K_1$ is the modified Bessel function of the first order. 
To find the edge-on disc central surface brightness, we should 
calculate $I_{0}^{edge-on}=\int_{-\infty}^{\infty} I(r,0)dr$ 
(designated as $\mu_\mathrm{0,d}$ if expressed in mag/arcsec$^2$). 

The bulge profile is given by the \ser\ law \citep{Sersic1968}:
%%%%%%%%%%%%%%%%%%%%%%%%%%%%%%%%%%%%%%%%%%%%%%%%%%%%%%%%%%%%%%%%%%%%%%
\begin{equation}
I(r) = 
  I_\mathrm{0,b} \, 
  e^{-\nu_n(r/r_\mathrm{e,b})^{1/n}}\, ,
\label{SBr}
\end{equation}
%%%%%%%%%%%%%%%%%%%%%%%%%%%%%%%%%%%%%%%%%%%%%%%%%%%%%%%%%%%%%%%%%%%%%%
where $r_\mathrm{e,b}$ is the effective radius, or the half-light 
radius, $I_\mathrm{0,b}$ is the central surface brightness (it can 
be reduced to the effective surface brightness $\mu_\mathrm{e,b}$ 
expressed in mag/arcsec$^2$), $n$ is the \ser\ index defining 
the shape of the profile, and the parameter $\nu_n$ depending on 
$n$ ensures that $r_\mathrm{e,b}$ is the half-light radius. 

In order to estimate the bulge-to-total luminosity ratio $B/T$, we 
can find the total luminosity of the bulge and the disc via integrating 
the equations~(\ref{form_DSB})~and~(\ref{SBr}) over the whole galaxy 
image.

{\sevensize DECA} also allows us to take into account the stellar disc 
truncation when needed. The axis ratio of the bulge $q_\mathrm{b}$ and its ellipticity 
parameter that controls the shape of the bulge (``boxy'' 
or ``discy'') can be set as free parameters.

%%%%%%%%%%%%%%%%%%%%%%%%%%%%%%%%%%%%%%%%%%%%%%%%%%%%%%%%%%%%%%%%%%%%%%
\subsection{Robustness of the {\sevensize DECA} decomposition}
\label{Simulations}
%%%%%%%%%%%%%%%%%%%%%%%%%%%%%%%%%%%%%%%%%%%%%%%%%%%%%%%%%%%%%%%%%%%%%%
As the main aim of this work is studying edge-on galaxies selected 
from SDSS, we tested {\sevensize DECA} on the sample of artificially 
created galaxy images that imitate SDSS images in the $i$ band 
with the typical for the EGIS galaxies signal-to-noise ratio. 
We generated a galaxy image as the sum of photometric disc and bulge 
components by using expressions~\eqref{form_DSB} and~\eqref{SBr},
respectively. We built a sample of 1000 artificial images with the 
uniform distributions of the parameters similar to those from 
\cite{Kregel+2002}. The boundary values for each parameter are taken 
from the physical background summarized in several works 
(e.g. \citealp{Gadotti2009}; MSR10). The disc parameters cover the 
following ranges:
\[
16\leq \mu_\mathrm{0,d}\leq 21\,\mathrm{mag\,arcsec}^{-2} \, ,
\]
\[
0.5\leq h\leq 10\,\mathrm{kpc} \, ,
\]
\[
0.1\leq z_0\leq 1.5\,\mathrm{kpc} \, .
\]

For bulge the ranges are:
\[
18\leq \mu_\mathrm{e,b}\leq 22\,\mathrm{mag\,arcsec}^{-2} \, ,
\]
\[
0.1\leq r_\mathrm{e,b}\leq 2\,\mathrm{kpc} \, ,
\]
\[
0.5\leq n\leq 5 \, ,
\]
\[
0.5\leq q_\mathrm{b}\leq 1.0 \, .
\]

The image scale varied randomly 
between 0.4 and 2 kpc/arcsec. The simulated images of galaxies were 
created using the Python wrapper 
{\sevensize SIGAL}\footnote{https://github.com/latrop/SIGAL} 
which operates the {\sevensize GALFIT} code to produce model images. 
%The images were simulated in such a way that they would resemble 
%SDSS images in the $i$ passband.
All created images were convolved 
with the Moffat PSF with $\mathrm{FWHM}=1.2$~arcsec. The pixel 
size is 0.396~arcsec/pix, the exposure time was set to 53.9~sec and the 
gain is 4\,e$^{-}$/ADU. We added gaussian readout noise and 
poison noise to each pixel of the image. All galaxy discs were 
truncated up to the distance of $4\,h$.

\begin{figure}
\centering
\includegraphics[width=2.5in, angle=0, clip=]{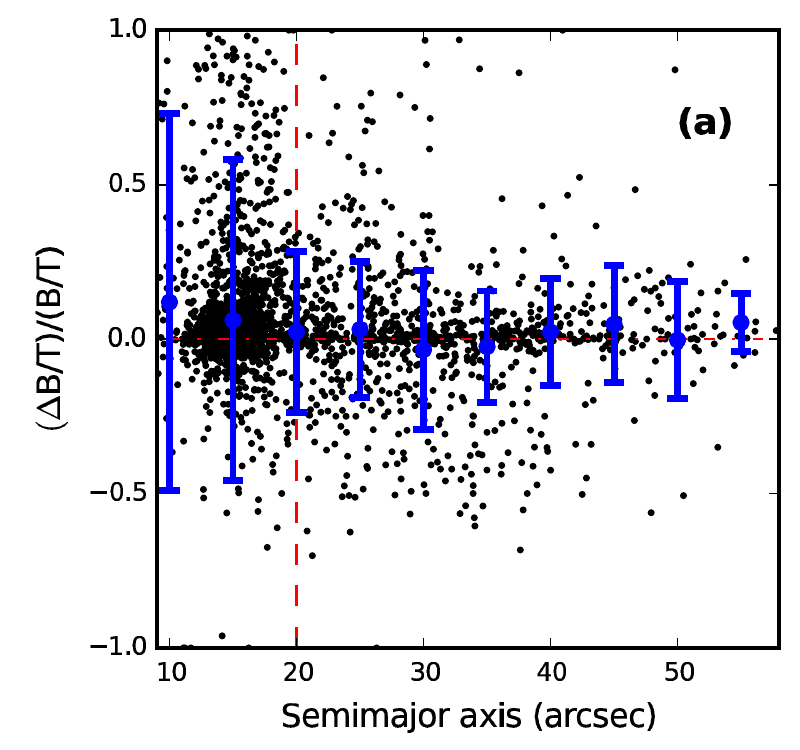}
\includegraphics[width=2.63in, angle=0, clip=]{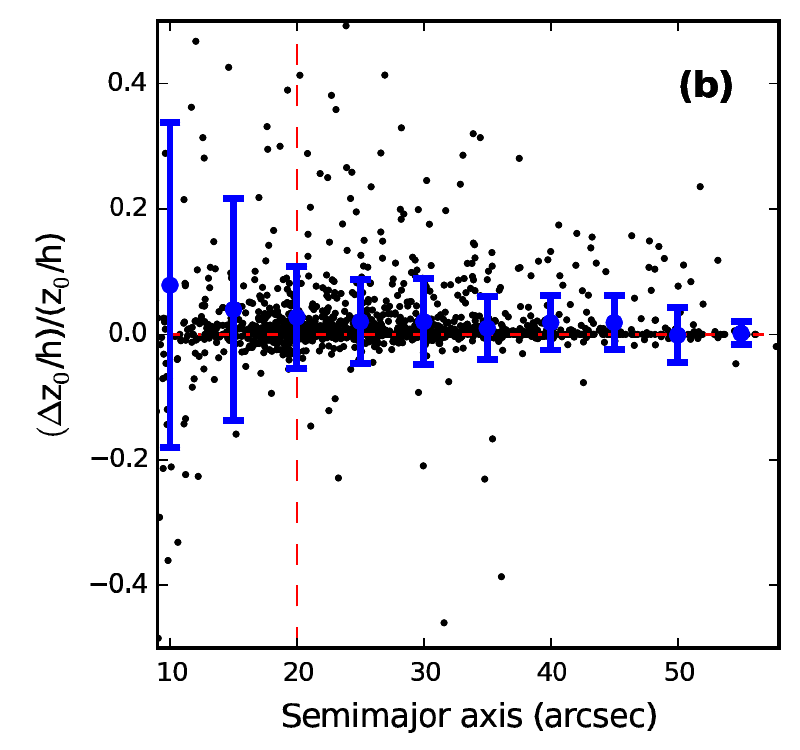}
\caption{ Results of simulations as a 
 function of the semi-major axis size of the artificially created edge-on 
 galaxy  images for the SDSS, $i$ band.
 (a) The distribution of 
 $(B/T-B/T_\mathrm{fit})/(B/T)$, where  $B/T$ is a known and 
 $B/T_\mathrm{fit}$ is a fitted bulge-to-total luminosity ratio.
 (b) The distribution of 
 $(z_0/h-z_0/h_\mathrm{fit})/(z_0/h)$, where  $z_0/h$ is a known and 
 $z_0/h_\mathrm{fit}$ is a fitted disc flattening.
 The limiting radius is estimated to be about $20$~arcsec 
 (designated as the vertical red dashed line), or 
 $D_\mathrm{min}^{i}=40$~arcsec. The ordinates for each plot were 
 averaged inside the bin of 5~arcsec (blue circles with the bars 
 representing the standard deviation of the parameters inside each 
 bin). The horizontal red dashed line  corresponds to the zero value 
 of the ordinate.}
\label{sim_sdss}
\end{figure}

The decomposition of the artificial galaxy images was then 
performed with the {\sevensize DECA} package. The robustness of 
the decomposition is shown in Fig.~\ref{sim_sdss} for the parameters 
$B/T$ and $z_0/h$ which are of great interest for us in this paper. We can see that for galaxies with a semi-major 
axis (being estimated as the truncation radius of the galaxy) 
larger than 20~arcsec 
$\langle(\Delta B/T) / (B/T)\rangle = 0.01\pm0.23$ and 
$\langle(\Delta z_0/h) / (z_0/h)\rangle = 0.017\pm0.06$ while 
for galaxies with the lesser semi-major axis 
$\langle(\Delta B/T) / (B/T)\rangle = 0.077\pm0.55$ and 
$\langle(\Delta z_0/h) / (z_0/h)\rangle = 0.05\pm0.18$. 
% The quality of the restoration of the parameter $z_0/h$ does 
% not depend on the galaxy semi-major axis size so critically as it 
% is for the parameter $B/T$ (compare plot (a) and (b)). 
Therefore we conclude that the minimum diameter of a galaxy with $B/T>0$ 
to be well-decomposed is $D_\mathrm{min}^{i}\approx40$~arcsec 
(hereafter we put the upper index to designate the passband). 
For galaxies with the diameter larger than $D_\mathrm{min}$ we 
may expect that the average errors for the fitted parameters $z_0/h$ 
and $B/T$ should be less than 10\% and 25\% of their true values 
respectively. 
Thus, the results of our test confirm the robustness of fitting performing 
by {\sevensize DECA} and its possibility for decomposing large sets 
of galaxies imaged by SDSS. 

%%%%%%%%%%%%%%%%%%%%%%%%%%%%%%%%%%%%%%%%%%%%%%%%%%%%%%%%%%%%%%%%%%%%%%
%%%% Fig_2
%%%%%%%%%%%%%%%%%%%%%%%%%%%%%%%%%%%%%%%%%%%%%%%%%%%%%%%%%%%%%%%%%%%%%%
%%%%%%%%%%%%%%%%%%%%%%%%%%%%%%%%%%%%%%%%%%%%%%%%%%%%%%%%%%%%%%%%%%%%%%

%%%%%%%%%%%%%%%%%%%%%%%%%%%%%%%%%%%%%%%%%%%%%%%%%%%%%%%%%%%%%%%%%%%%%%
\section[]{The samples and the decomposition}
\label{s_samples}
%%%%%%%%%%%%%%%%%%%%%%%%%%%%%%%%%%%%%%%%%%%%%%%%%%%%%%%%%%%%%%%%%%%%%%
Our further analysis is based on the study of several samples of 
galaxies viewed almost perfectly edge-on ($i\geq86^{\circ}$) that were observed 
in different passbands. Although we focus our study on the sample 
of galaxies selected from SDSS, we also use the results of several 
investigations in near- and mid-infrared bands.

The Sloan Digital Sky Survey \citep{Aihara+2011} is an ambitious 
project to investigate the Universe in five optical passbands. 
Because the dust extinction along the major axis of the disc is 
substantially high in these bands, the vast majority of edge-on 
galaxy images demonstrate very well seen dust lanes. Hence, the proper 
model fitting of such galaxies should be carried out with taking into 
account the radiative transfer approach 
\citep[e.g.][]{Xilouris+1999} 
which can be done for galaxies with sufficient spatial resolution 
(for nearby edge-on galaxies if we use SDSS) and takes significant computational time. 
Nevertheless, as we will show later, some galaxies 
(of both early and late types) do not demonstrate sharp dust lanes 
in optics, and thus they can be decomposed using standard fitting 
packages described in Section~\ref{Techniques}. 
Galaxies with dust traces (having no sharp dust lanes) can be also 
decomposed using masking of these regions of high dust extinction. 

Since in the near-infrared ($JHK$) and mid-infrared Spitzer bands $W1$ 
(3.6 $\mu$m) and $W2$ (4.5 $\mu$m) the flux is largely dominated by 
the stellar radiation rather than gas and dust emission \citep{whaley09}, 
and the dust extinction here is much 
less than in optics, it is crucial to investigate the structure of 
the stellar discs and bulges in these bands. Because of these aspects, 
the use of 2MASS \citep{Skrutskie+2006} and S$^4$G 
\citep{sheth+2010} is suitable for our aims.

In order to select edge-on galaxies from existing catalogues and 
databases, we will use a selection criterion based on the galaxy 
apparent diameter at a given isophotal level.
%(excepting S$^4$G where the selection is based primarily on the basis of the pre-selected galaxies with the performed decomposition).
% of 25 mag~arcsec$^{-2}$ in the $B$ band $d_{25}(B)$ taken from the HyperLeda. 
This allows us to select galaxies in the 
same manner. As was noticed in Section~\ref{Simulations}, the value 
of the limiting apparent galaxy diameter for each sample 
should be chosen so as to exclude galaxies with poor 
bulge/disc decomposition. 
In Section~\ref{Simulations} we found that 
$D_\mathrm{min}^{i}=40$~arcsec if we consider galaxy images in SDSS. 
Supposing that the main factor that affects the quality of 
bulge/disc decomposition is the image resolution and PSF
(as reflected by the pixel size in arcsec), we can estimate the minimum diameter 
of 2MASS and S$^4$G galaxies to be well decomposed by scaling our estimates 
performed for SDSS case.   
The pixel size in 2MASS is 2.5 times larger than in SDSS, 
hence the limiting apparent diameter should be 
$D_\mathrm{min}^{Ks}=100$~arcsec. 
In the case of S$^4$G where the pixel size is 0.75~pixel/arcsec, 
$D_\mathrm{min}^{3.6}=75$~arcsec.  

%For that, we carried out simulations of 2MASS and SDSS galaxy images. We created artificial images of edge-on galaxies with a wide range of parameters taking into account the pixel size, the point spread function PSF and gaussian and poison noise of the CCD images of these surveys. In each case we also noted that a galaxy should have a sufficiently resolved bulge (with the effective radius $r_\mathrm{e,b}>0.4~\mathrm{FWHM}$, see \citealp{Gadotti+2010}) to be properly decomposed onto several components. 

In our study we consider only two main galactic components: 
the bulge and the disc, for simplicity. Possible 
influence of the multi-component structure of edge-on galaxies on 
the main results of this work will be discussed in 
Section~\ref{model_dependence}. 

The galaxy morphological type ($Type$) and the colour $B-I$ for 
each galaxy were taken from the HyperLeda database 
\citep{Paturel+2003}. The $B$ and $I$ bands are close to the $g$ 
and $i$ bands, respectively, therefore we opt to use the colour 
$B-I$ for 2MASS and S$^4$G galaxies along with the colour $g-i$ 
for SDSS galaxies.  

Each galaxy from the samples was visually inspected whether its shape 
resembles the shape of edge-on galaxies. Images in different 
passbands available in the NASA/IPAC Extragalactic Database (NED)
were analyzed to check the presence of 
a dust lane along the edge-on disc. For all galaxies of the samples 
we retrieved inclination angle $i$ from the HyperLeda and found 
the mean value of all galaxies studied 
$\langle i \rangle=88.\degr0\pm3.\degr5$. In 
Appendix~\ref{Appendix} we describe how we estimated the 
inclination angle in an independent way.

\begin{figure*}
\includegraphics[width=17.0cm, angle=0, clip=]{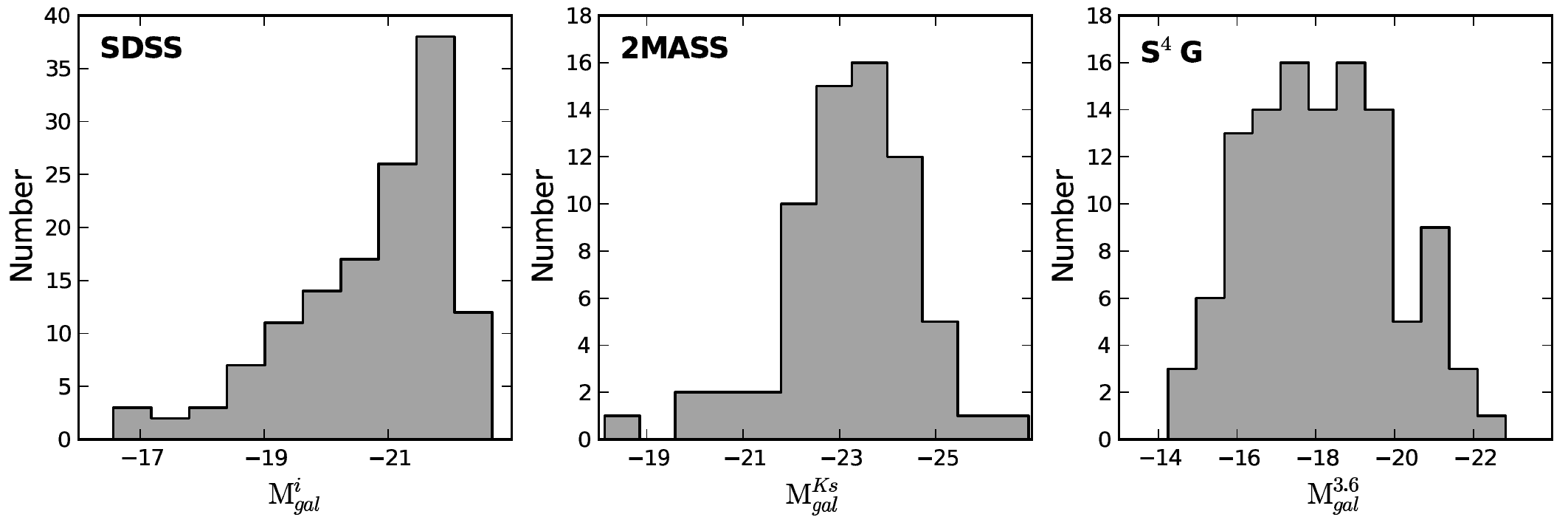}
\caption{The distributions of the total luminosity for the galaxies from SDSS 
 ($i$ band), 2MASS ($K_\mathrm{s}$ band) and S$^{4}$G (3.6~$\mu$m) samples.}
\label{distributions}
\end{figure*}

Fig.~\ref{distributions} demonstrates the distribution of the total 
galaxy luminosities in our samples. We ensure that the galaxies used 
for the analysis are non-dwarf and luminous. 
Detailed description of the samples is given below. 

%%%%%%%%%%%%%%%%%%%%%%%%%%%%%%%%%%%%%%%%%%%%%%%%%%%%%%%%%%%%%%%%%%%%%%
%%%% Fig_3
%%%%%%%%%%%%%%%%%%%%%%%%%%%%%%%%%%%%%%%%%%%%%%%%%%%%%%%%%%%%%%%%%%%%%%
%%%%%%%%%%%%%%%%%%%%%%%%%%%%%%%%%%%%%%%%%%%%%%%%%%%%%%%%%%%%%%%%%%%%%%

%%%%%%%%%%%%%%%%%%%%%%%%%%%%%%%%%%%%%%%%%%%%%%%%%%%%%%%%%%%%%%%%%%%%%%
\subsection[]{The SDSS sample}
\label{SDSS}
%%%%%%%%%%%%%%%%%%%%%%%%%%%%%%%%%%%%%%%%%%%%%%%%%%%%%%%%%%%%%%%%%%%%%%
In order to select comparatively large edge-on galaxies from
SDSS, we used the EGIS catalogue which is described in detail by 
BKM14. This is the largest catalogue of verified edge-on galaxies at the 
moment which comprises 5747 genuine edge-on galaxies pre-selected 
automatically from the SDSS Data Release 8 and then 
visually inspected. This catalogue is complete for all galaxies 
with the major axis $D^{i}$ larger than $28$~arcsec (see BKM14) and 
comprises disc galaxies of all morphological types (Sa--Sd-Irr), which 
is important for studying galaxies of different bulge contribution.

In Section~\ref{Simulations} we concluded that the minimum diameter of 
galaxies for keeping our decomposition robust is 
$40$~arcsec. Therefore, from the EGIS catalogue we selected objects 
with $D^{i}>40$~arcsec  and excluded galaxies with 
irregular structure of discs and with various contaminants as bright 
overlapping stars or galaxies. This resulted in the initial sample 
of 2027 objects. After a thorough visual inspection and preliminary 
decomposition of each galaxy image in the $i$ band, we selected 
145 galaxies and then estimated their inclination angle using 
two methods described in Appendix~\ref{Appendix}. We applied the 
first method to the galaxies without dust traces and the second 
method to the galaxies with visible dust manifestations. After 
that, we removed 4 galaxies with $i<86\degr$. Therefore, the 
final sample includes 82 galaxies without any presence of dust 
lanes and 59 galaxies with slightly visible dust lanes. The 
mean inclination angle for the galaxies in our sample is $\langle i \rangle=88.\degr2$. 
An evidence that a galaxy has no strong dust attenuation came from the 
inspection of residual images (`galaxy'$-$`model') and browsing
of the vertical and radial photometric profiles. Since no dust traces 
were found in their profiles, we may hope that structural 
decomposition of these galaxies based on the model of transparent 
disc is more robust than for the rest of the sample. We will call 
this subsample of 82 `dust-free' galaxies the ``reference subsample". 
% Possible effects of the internal extinction will be discussed 
% in Section~\ref{model_dependence}.
The selected sample consists 
of 27 early-type ($T\leq0$) and 114 late-type galaxies ($T>0$). 
The redshifts retrieved 
from the SDSS Data Release 9 prove that the galaxies of the sample 
are nearby ($z<0.05$, $\langle z\rangle=0.02$).

\begin{figure*}
% \centering
\begin{minipage}[h]{0.47\linewidth}
\center{\includegraphics[width=2.2in, angle=0, clip=]{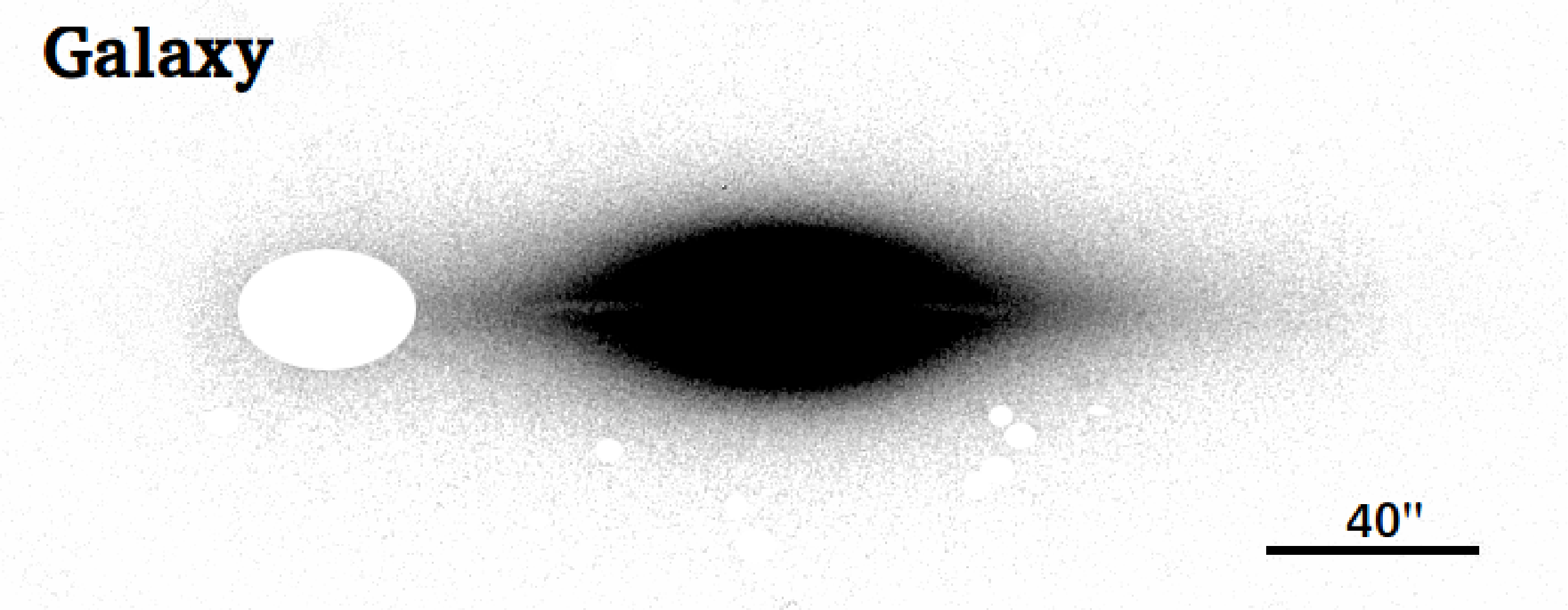}}
\vfill

\center{\includegraphics[width=2.2in, angle=0, clip=]{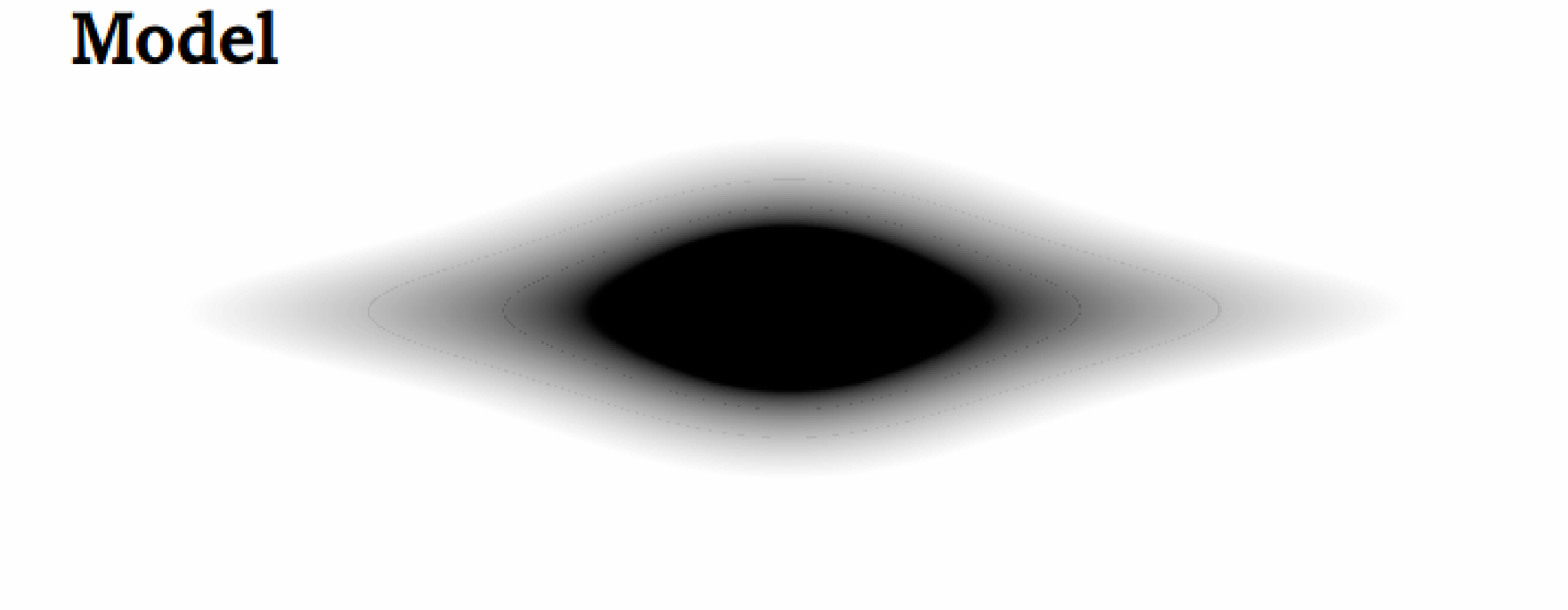}}
\vfill

\center{\includegraphics[width=2.2in, angle=0, clip=]{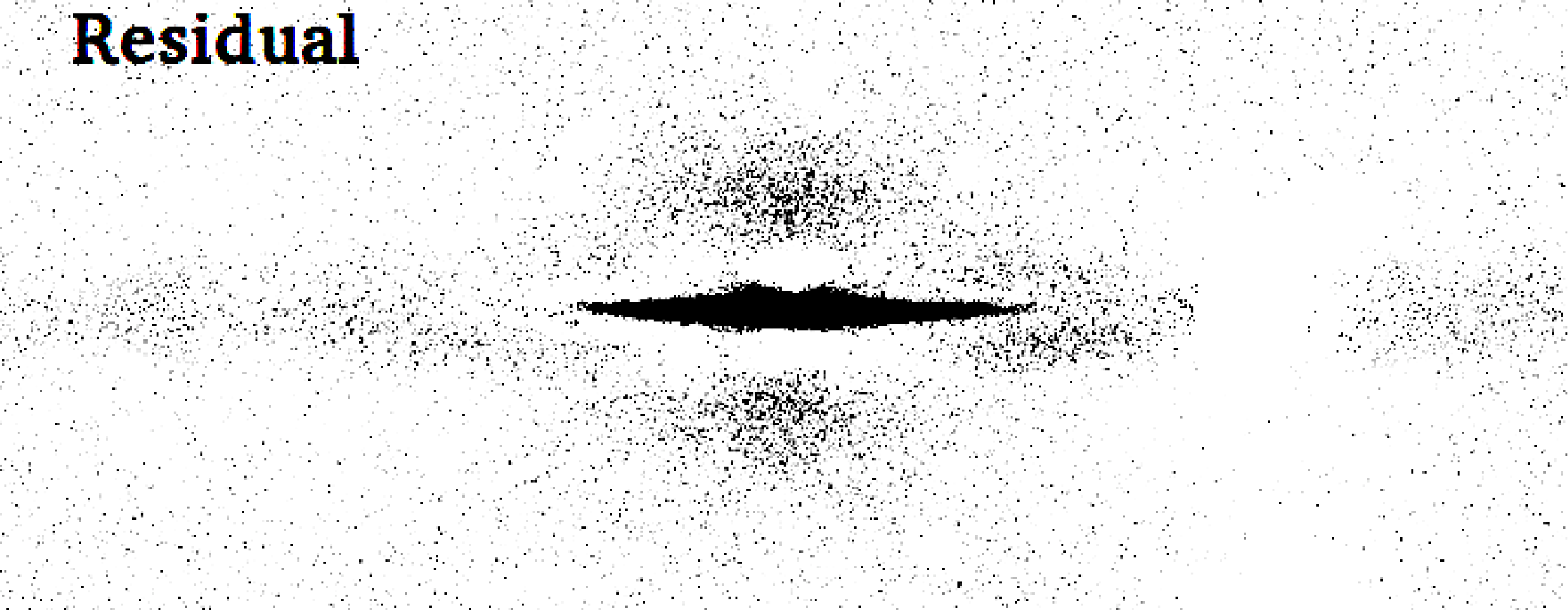}}
\end{minipage}
\hfill
\begin{minipage}[h]{0.47\linewidth}
\center{\includegraphics[width=3.2in, angle=0, clip=]{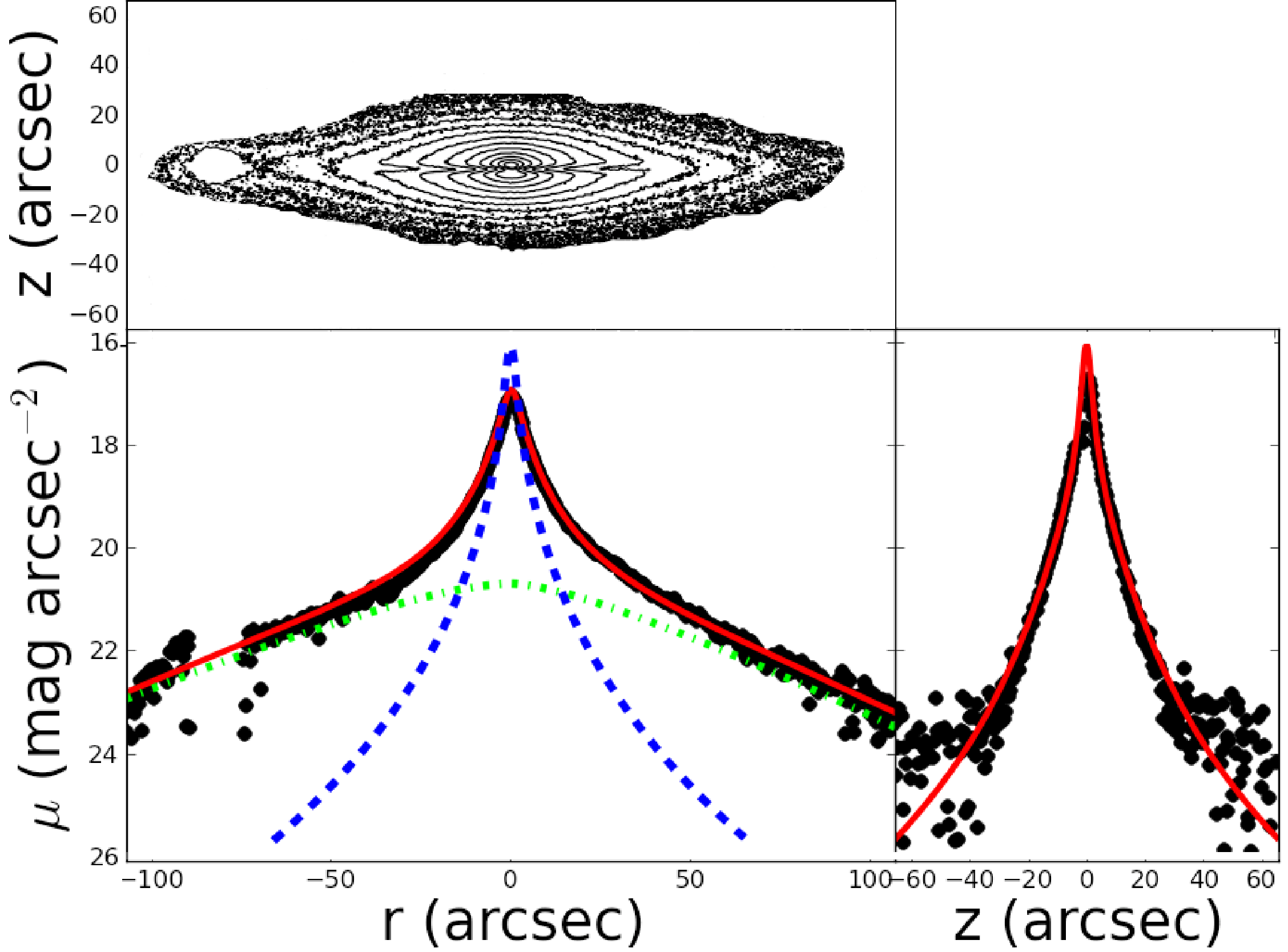}}
\end{minipage}
\caption{The decomposition of the S0/a edge-on galaxy NGC~1032 using 
 the {\sevensize DECA} code (SDSS, $i$ band). Left panel shows the galaxy 
 image (top), the model (middle) and the 
 residual image (`galaxy'$-$`model', bottom). Right panel represents 
 the isophote map of the galaxy down to the 22.5~mag\,arcsec$^{-2}$ surface brightness 
 (top), the photometric cut along the major axis (bottom left) 
 and the photometric cut along the minor axis (bottom right). 
 The black dots refer to the galaxy, the red solid line represents 
 the model, the blue dashed line represents the bulge, and the disc 
 is shown by the green dots. The dust lane was masked during the 
 decomposition. Main derived parameters of the decomposition are 
 $B/T=0.70$, $z_0/h=0.30$.}
% Right images from top to bottom are the galaxy, 
%   the model, and the residual image. 
%   
  %The derived parameters of the disc are 
  %$\mu_{0,d}$=20.56\,mag\,arcsec$^{-2}$, 
  %$h$=35.1\,arcsec, $z_\mathrm{0}$=10.7\,arcsec; 
  %for the bulge: $\mu_\mathrm{e,b}$=19.66\,mag\,arcsec$^{-2}$, 
  %$r_\mathrm{e,b}$=14.8\,arcsec, $n$=3.13, and the apparent bulge axis ratio $q_\mathrm{b}$=0.6. }
\label{ex_dec}
\end{figure*}

The 2D decomposition of the galaxy images from the sample of all
141 objects was performed using the {\sevensize DECA} code. 
The $i$ band images were downloaded from the SDSS Data Release 9 
\citep{Ahn+2012}. PSF (prefix ``psField'') images were retrieved 
from the SDSS Data Archive Server (DAS). If a prominent dust lane was 
seen in the galaxy image (as for NGC~1032 in 
Fig.~\ref{ex_dec}), it was masked during the decomposition.

%%%%%%%%%%%%%%%%%%%%%%%%%%%%%%%%%%%%%%%%%%%%%%%%%%%%%%%%%%%%%%%%%%%%%%
%%%% Fig_
%%%%%%%%%%%%%%%%%%%%%%%%%%%%%%%%%%%%%%%%%%%%%%%%%%%%%%%%%%%%%%%%%%%%%%
%%%%%%%%%%%%%%%%%%%%%%%%%%%%%%%%%%%%%%%%%%%%%%%%%%%%%%%%%%%%%%%%%%%%%%

\begin{figure*}
\includegraphics[width=18.0cm, angle=0, clip=]{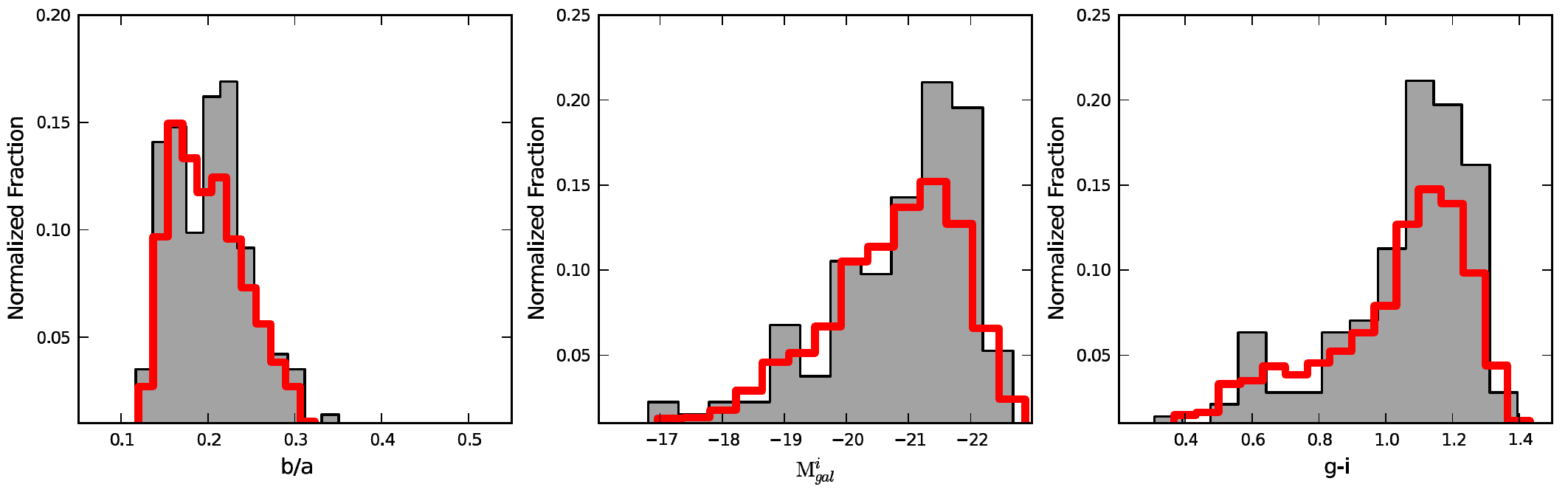}
\caption{The distributions of the apparent axis ratio, total luminosity 
 and colour for the selected SDSS sample (gray) and the EGIS 
 catalogue (red solid line).}
\label{SDSS_full_my}
\end{figure*}

We checked our sample for completeness and found that the subsample 
of 98 galaxies with $D^{i}>48$~arcsec is complete while 
the whole sample is incomplete according to the $V/V_\mathrm{max}$ 
test \citep{Thuan+1979}. In Fig.~\ref{SDSS_full_my}, left panel, we can see that the 
distribution of the apparent axis ratio in the sample (gray color) and 
in the whole EGIS catalogue (red solid line) are very similar. 
We performed the Kolmogorov-Smirnov test to verify the null 
hypothesis that these two samples are drawn from the same population. 
We found that we cannot reject the null hypothesis at 
a 5$\%$ or lower $\alpha$ since the $p$-value is high and equals
to 0.88. If the distributions of galaxy luminosity 
(Fig.~\ref{SDSS_full_my}, middle plot) and colour are considered,
(Fig.~\ref{SDSS_full_my}, right plot) we can also see that these 
distributions look similar for both samples. However, according 
to the Kolmogorov-Smirnov test there is no sufficient evidence 
to reject the null hypothesis (for the $M_\mathrm{gal}^{i}$ 
distribution: $\alpha=0.105$, $p$-value$=0.12$; for the $g-i$ 
distribution: $\alpha=0.09$, $p$-value$=0.14$). 
Since the apparent axis ratio, galaxy total luminosity and colour 
are the main photometrical characteristics of the observed objects, 
the distributions of these parameters for the sample of 141 galaxies 
and the EGIS catalogue are similar, we may conclude that our sample
well represents the whole EGIS catalogue.

%The mean galaxy inclination retrieved from HyperLeda for the whole sample of 145 galaxies 
%is $i=88.3^{\circ}\pm3.0^{\circ}$. Inclination estimations 
%in HyperLeda are rather rough, but statistically they show that 
%this sample consists of truly edge-on galaxies. Moreover, `discy' isophotes 
%and sharp tips on the periphery of the galaxies is another evidence 
%that these galaxies are viewed edge-on. We ????? 

% The sample comprises both early ($N$(S0--Sa)=59) and late spirals ($N$(Sb--Sd)=85). All the galxies are nearby ($z<0.05$).
% The whole sample is incomplete, but the subsample of 99 galaxies
% with $d_{25}(B)>48\arcsec$ is complete with $V/V_\mathrm{max}=0.502\pm 0.03$ \citep{Thuan+1979}. 

%Galaxy images in the $i$ band were downloaded from the SDSS DR9 \citep{Ahn+2012},
%which has much better sky subtraction than the previous releases. All 
%stars, background and underground objects were masked. 

% From the sample of 192 galaxies we selected 55 galaxies with 
% $d_{25}(B)>60$~arcsec. The models for these galaxies look fine, 
% with minimum of $\chi^2$. The outer isophotes of these galaxies 
% are needle-like which confirms our suggestion that these are true 
% edge-on galaxies. We will call these group of galaxies reliable 
% edge-ons. All other galaxies seem to be little inclined which 
% follows from their ellipse shape of isophotes or have the diameter 
% $d_{25}(B)\leq60$~arcsec  (see for comparison Fig. ..).

%%%%%%%%%%%%%%%%%%%%%%%%%%%%%%%%%%%%%%%%%%%%%%%%%%%%%%%%%%%%%%%%%%%%%%
%%%% Fig_5
%%%%%%%%%%%%%%%%%%%%%%%%%%%%%%%%%%%%%%%%%%%%%%%%%%%%%%%%%%%%%%%%%%%%%%
%%%%%%%%%%%%%%%%%%%%%%%%%%%%%%%%%%%%%%%%%%%%%%%%%%%%%%%%%%%%%%%%%%%%%%

%%%%%%%%%%%%%%%%%%%%%%%%%%%%%%%%%%%%%%%%%%%%%%%%%%%%%%%%%%%%%%%%%%%%%%
\subsection[]{The 2MASS sample}
\label{2MASS}
%%%%%%%%%%%%%%%%%%%%%%%%%%%%%%%%%%%%%%%%%%%%%%%%%%%%%%%%%%%%%%%%%%%%%%
% This sample and the decomposition technique are described in MSR10. As a 
% source of objects we used the 2MASS-selected Flat Galaxy Catalog 
% (2MFGC, \citealp{Mitronova+2003}). The fits-images of 175 galaxies 
% in the $K_\mathrm{s}$ band were taken from the 2MASS archive. From this sample 
% we also selected large galaxies with 
% $d_{25}(B)>120$~arcsec, 
% for which the mid-plane dust layer in the optical bands (analyzing galaxy images 
% from DSS and SDSS) overlaps the central horizontal plane of the 
% galaxy's stellar image. The total number of these genuine edge-on galaxies is 
% 68, and we analyzed them separately. 
In order to create another sample of genuine edge-on galaxies, we 
used a sample described in detail in MSR10. The 2MASS sample from 
MSR10 contains galaxies with a wide range of bulge contribution: 
from galaxies with prominent bulges to bulgeless galaxies. To select 
the galaxies, we used the 2MASS-selected Flat Galaxy Catalog 
(2MFGC, \citealp{Mitronova+2003}) and the Revised Flat Galaxy 
Catalog (RFGC) by \cite{Karachentsev+1999}. 
% The last catalogue contains mainly late-type edge-on galaxies, and we selected the largest edge-on galaxies from there.
Galaxies were selected from the 2MFGC according to the following 
criteria: the axial ratio $sba>0.2$, the $K_\mathrm{s}$ band Kron 
radius $R^{K_\mathrm{s}}\geq30$~arcsec and the concentration index 
(which is connected to the bulge-to-disc luminosity ratio) 
$IC>2.0$. The interacting, peculiar and non-edge-on galaxies 
identified visually and according to the information found in 
HyperLeda, were 
removed from our consideration.
The final sample consists of 175 galaxies in the 
$K_\mathrm{s}$ band, with images taken from the 2MASS archive. 
All the galaxies were decomposed into a bulge and a disc using the 
{\sevensize BUDDA} code \citep{desouza+2004} (see several examples of 
decomposition in MSR10).

We found that about one-third out of the sample of 175 galaxies have 
inclinations less than $86^{\circ}$ as estimated using 
the first method described in Appendix~\ref{Appendix}.  
We removed these galaxies, as well as the galaxies with the 
diameter smaller than $D_\mathrm{min}^{K_\mathrm{s}}=100$~arcsec. 
The resulting sample comprises 66 large genuine edge-on galaxies. 

\begin{figure*}
\includegraphics[width=18.0cm, angle=0, clip=]{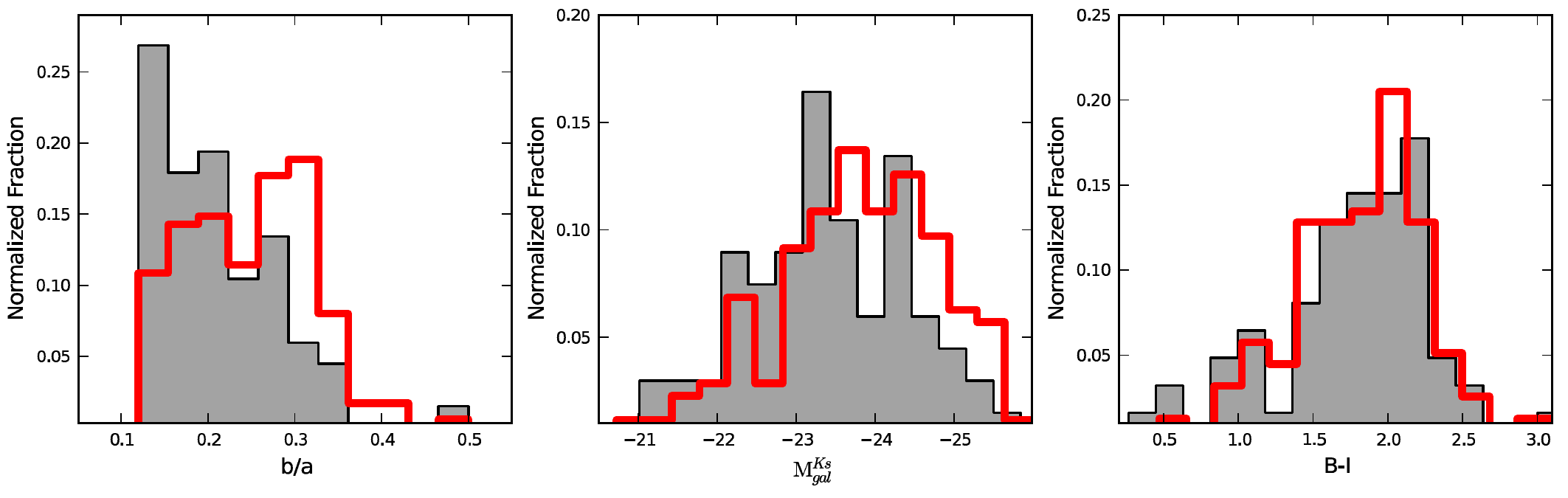}
\caption{The distributions of the apparent axis ratio, total luminosity 
 and colour for the selected 2MASS sample (gray) and the sample 
 from MSR10 (red solid line).}
\label{2MASS_full_my}
\end{figure*}

According to the $V/V_\mathrm{max}$ test our 
final sample is incomplete and thus may suffer selection effects. 
In the left panel in Fig.~\ref{2MASS_full_my} we can see that the 
distribution of the final sample (gray color) and the distribution of 
the whole 2MASS sample over the apparent axis ratio differ since 
we excluded non edge-on galaxies from subsequent analysis -- 
these galaxies form a peak at $b/a=0.32$ (the red line distribution). 
The other distributions (middle and right panels) do not show 
significant differences between the initial sample of 175 galaxies 
and the final one. The selected sample comprises 10 early-type and 
56 late-type galaxies. All the galaxies are nearby, with $z<0.035$ 
(average redshift $\langle z\rangle=0.007$); the mean inclination angle is $\langle i \rangle=88.\degr0$.
%for which the mid-plane dust layer in the optical bands (analyzing galaxy images from DSS and SDSS) overlaps the central horizontal plane of the galaxy's stellar image. The total number of these nearby ($z<0.02$), genuine edge-on galaxies is 41. This robust subsample will be analyzed separately in contrast with the complete subsample of 92 galaxies. 

%%%%%%%%%%%%%%%%%%%%%%%%%%%%%%%%%%%%%%%%%%%%%%%%%%%%%%%%%%%%%%%%%%%%%%
%%%% Fig_6
%%%%%%%%%%%%%%%%%%%%%%%%%%%%%%%%%%%%%%%%%%%%%%%%%%%%%%%%%%%%%%%%%%%%%%
%%%%%%%%%%%%%%%%%%%%%%%%%%%%%%%%%%%%%%%%%%%%%%%%%%%%%%%%%%%%%%%%%%%%%%

%%%%%%%%%%%%%%%%%%%%%%%%%%%%%%%%%%%%%%%%%%%%%%%%%%%%%%%%%%%%%%%%%%%%%%
\subsection[]
{The S$^{4}$G sample}
\label{S4G}
%%%%%%%%%%%%%%%%%%%%%%%%%%%%%%%%%%%%%%%%%%%%%%%%%%%%%%%%%%%%%%%%%%%%%%
The S$^{4}$G catalogue consists of 2352 nearby (with distance less 
than $40~$Mpc) galaxies at 3.6 and 4.5$~\mu$m, and it is a valuable 
source of edge-on galaxies. Using the S$^{4}$G Galfit Models Home 
Page\footnote{The catalogue of galactic structural parameters can be found at 
\textit{http://www.oulu.fi/astronomy/S4G\_PIPELINE4/s4g\_p4\_table8.dat}}, 
we retrieved the best decomposition 
models with the edge-on disc included in 362 galaxies 
(at 3.6$\mu$m). However, the visual inspection of 
 the Spitzer, DSS and SDSS
images of these 
galaxies showed that some 
galaxies are not quite edge-on since they have round elliptical 
isophotes, or visible spiral arms, or rings), therefore we did not 
classify these galaxies as edge-on (for example, NGC~5984 and 
NGC~678). Otherwise, we would risk to retrieve false thicker stellar discs, 
which can significantly mislead us in studying the dependence between 
the disc flattening and the galaxy morphological type.

From the pre-selected sample of edge-on galaxies in the S$^4$G model 
catalogue we selected galaxies consisted either of the disc alone 
(late types), or of the disc and the bulge (early and intermediate 
types\footnote{31 galaxies were fitted with the `bulge+disc' model}). 
Galaxies with detected double-exponential discs are not considered 
in this work. Then, we estimated inclinations of the selected 
galaxies using the first method from Appendix~\ref{Appendix} and 
removed objects with $i<86\degr$. Also, we removed non-bulgeless 
galaxies with smaller than minimum diameter at the outer isophote of 
$W1=25.5$~mag\,arcsec$^{-2}$ $D_{25.5}^{3.6}<D_\mathrm{min}^{3.6}$, 
where $D_\mathrm{min}^{3.6}$ was estimated to be $75$~arcsec.

The total number of edge-on galaxies of all types 
selected from the S$^4$G model catalogue is 99, with the mean 
inclination angle $\langle i \rangle=87.\degr9$. 
The sample mostly consists of the late-type galaxies (8 early-type 
and 91 late-type galaxies). We did not reject galaxies of the 
latest morphological types (23 irregular galaxies, with poorly 
defined structure) though we assume that their edge-on orientation 
determination is rather precarious. We make some comments on this 
issue in the further sections. 

\begin{figure*}
\includegraphics[width=18.0cm, angle=0, clip=]{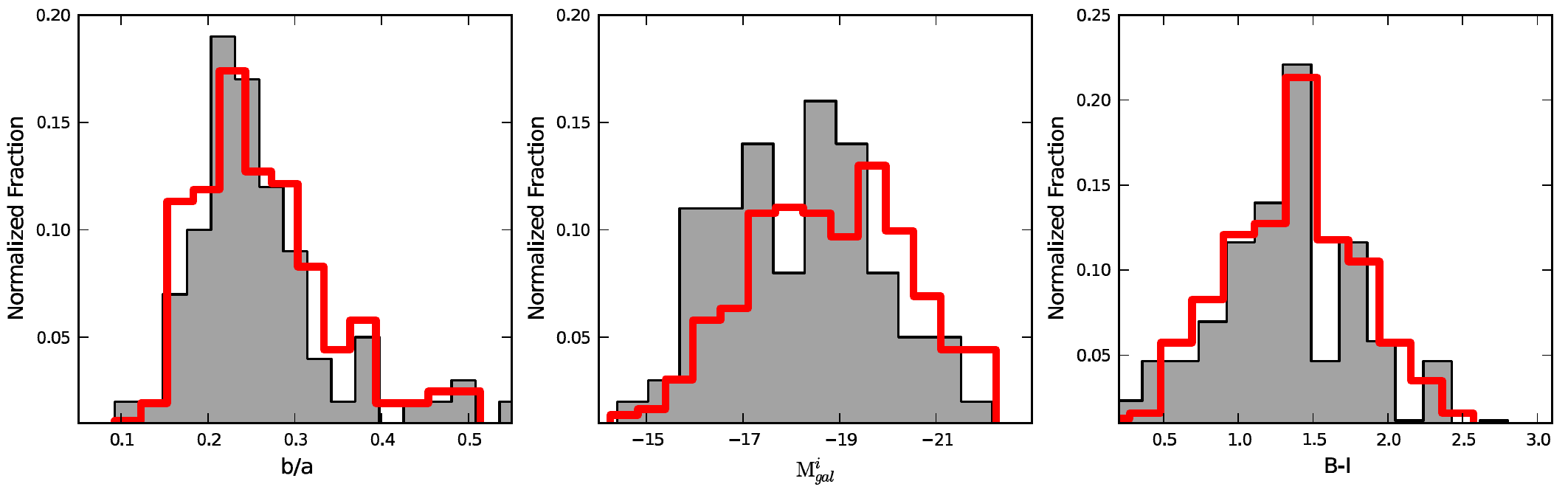}
\caption{The distributions of the apparent axis ratio, total luminosity 
 and colour for the selected S$^4$G sample (gray) and the whole 
 sample of 362 edge-on galaxies in the S$^{4}$G catalogue (red 
 solid line).}
\label{S4G_full_my}
\end{figure*}

In Fig.~\ref{S4G_full_my} we show the distributions for the 
selected sample (gray color) and the whole sample of 362 galaxies 
decomposed with edge-on discs in the S$^{4}$G catalogue 
(the red line distribution). It is well seen that all present distributions 
are very similar. However, the distribution of
$M_\mathrm{gal}^{i}$ shows that our S$^4$G sample comprises larger 
number of low-luminosity galaxies and less number of high 
luminosity galaxies in the comparison with the whole sample of edge-on 
galaxies in the S$^4$G catalogue.

%%%%%%%%%%%%%%%%%%%%%%%%%%%%%%%%%%%%%%%%%%%%%%%%%%%%%%%%%%%%%%%%%%%%%%
%%%% Fig_7
%%%%%%%%%%%%%%%%%%%%%%%%%%%%%%%%%%%%%%%%%%%%%%%%%%%%%%%%%%%%%%%%%%%%%%
%%%%%%%%%%%%%%%%%%%%%%%%%%%%%%%%%%%%%%%%%%%%%%%%%%%%%%%%%%%%%%%%%%%%%%

\begin{figure}
\includegraphics[width=8.0cm, angle=0, clip=]{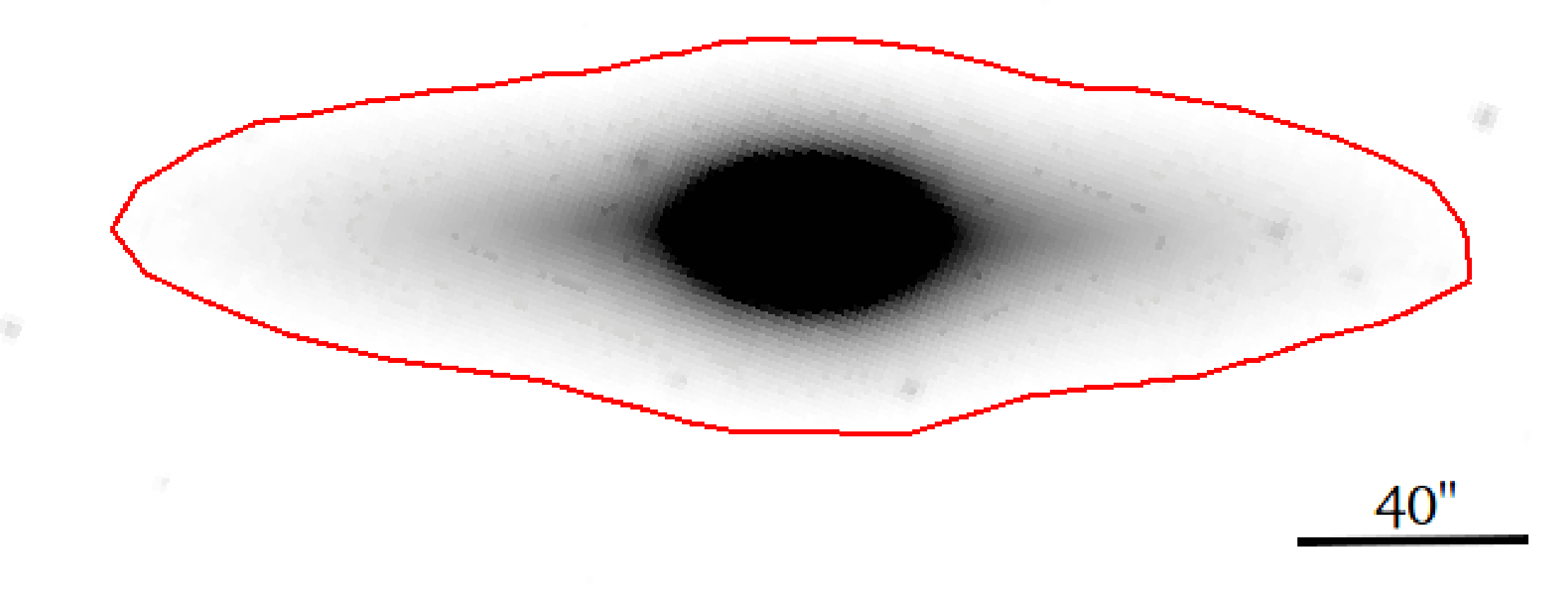}
\includegraphics[width=8.0cm, angle=0, clip=]{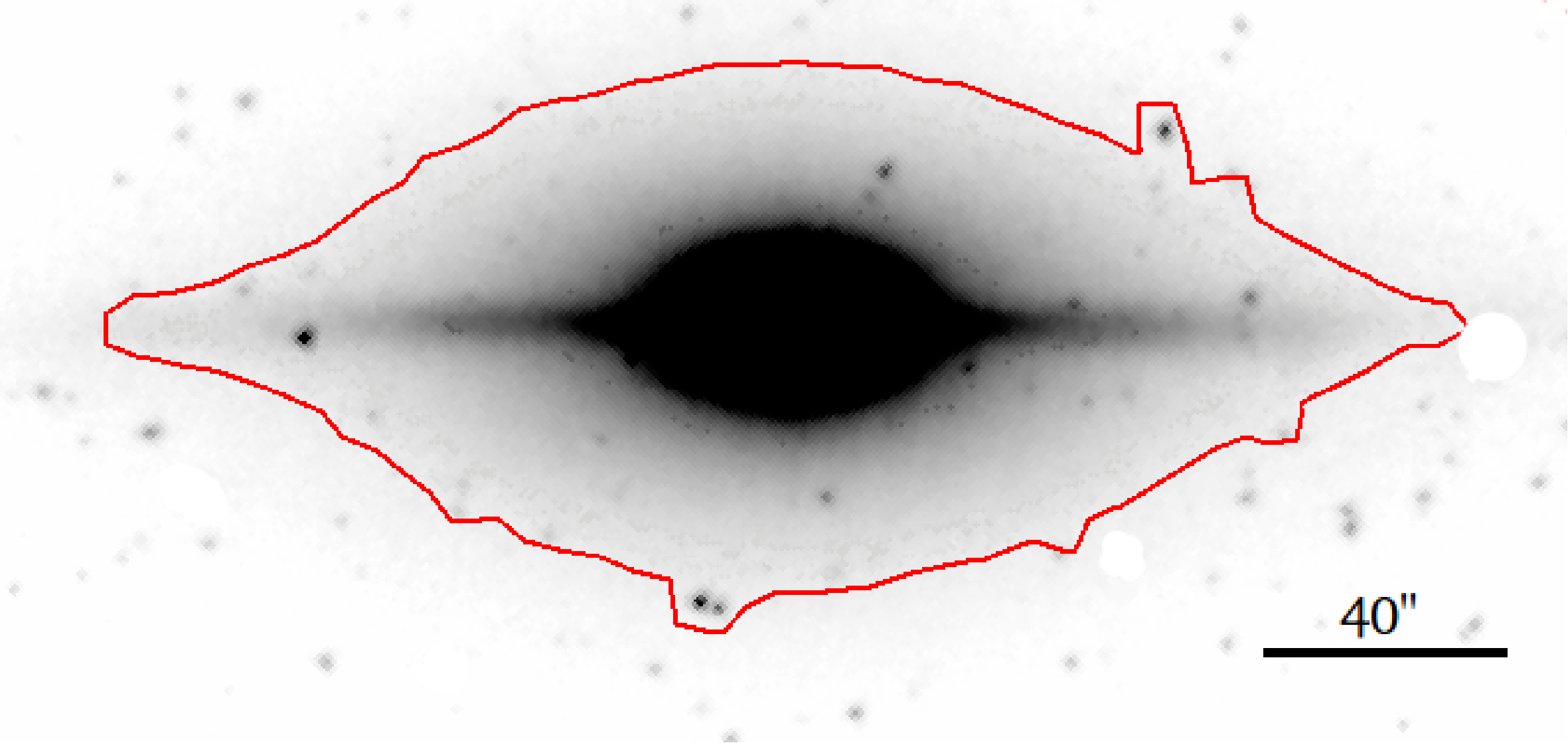}
\caption{Examples of early-type galaxies from the S$^4$G database 
 (W1 band): NGC~1596 (S0, $z_0/h=0.34$, $B/T=0.64$, top) 
 and NGC~7814 (Sab, $z_0/h=0.15$, $B/T=0.90$, bottom).  
 The disc thicknesses and bulge-to-total luminosity ratios are taken from S$^4$G Galfit Models Home Page. 
 The red contours refer to the outer isophote of 
 W1=25.5~mag\,arcsec$^{-2}$.}
\label{galaxies_ex_s4g}
\end{figure}

Although our S$^4$G sample is significantly incomplete according to the 
$V/V_\mathrm{max}$ test and different morphological types
are not statistically representative in the sample, this is the only sample of 
genuine edge-on galaxies observed in the mid-infrared bands 
(see examples of the edge-on galaxies from the 
created S$^4$G sample in~Fig.~\ref{galaxies_ex_s4g}), thus 
we consider this sample carefully in the current work. 

%%%%%%%%%%%%%%%%%%%%%%%%%%%%%%%%%%%%%%%%%%%%%%%%%%%%%%%%%%%%%%%%%%%%%%
%%%% Fig_8
%%%%%%%%%%%%%%%%%%%%%%%%%%%%%%%%%%%%%%%%%%%%%%%%%%%%%%%%%%%%%%%%%%%%%%
%%%%%%%%%%%%%%%%%%%%%%%%%%%%%%%%%%%%%%%%%%%%%%%%%%%%%%%%%%%%%%%%%%%%%%

%%%%%%%%%%%%%%%%%%%%%%%%%%%%%%%%%%%%%%%%%%%%%%%%%%%%%%%%%%%%%%%%%%%%%%
\section{Results}
\label{Results}
%%%%%%%%%%%%%%%%%%%%%%%%%%%%%%%%%%%%%%%%%%%%%%%%%%%%%%%%%%%%%%%%%%%%%%

%%%%%%%%%%%%%%%%%%%%%%%%%%%%%%%%%%%%%%%%%%%%%%%%%%%%%%%%%%%%%%%%%%%%%%
\subsection[]{The disc thickness versus the morphological type}
\label{res1}
%%%%%%%%%%%%%%%%%%%%%%%%%%%%%%%%%%%%%%%%%%%%%%%%%%%%%%%%%%%%%%%%%%%%%%
First, let us consider the dependence of the disc relative 
thickness $z_0/h$ on the galaxy type. In fig.~14 from BKM14 one can 
see the dependence between the stellar disc thickness and the type
derived from the 1D analysis. This dependence is similar to that found 
in \citet{deGrijs1998} also based on the analysis of 1D 
photometric profiles. At the same time, the 3D analysis for the same 
sample (see fig.~15 in BKM14) does not allow us to conclude that this 
dependence truly exists. It only shows that the late-type edge-on 
galaxies (determined as galaxies without bulge) are thinner 
than others, on average. Thus, we can conclude that a more advanced 
3D method of decomposition erases the thickness-type dependence. 

It is worth noticing that morphological types in BKM14 were 
estimated automatically for each galaxy using main photometrical 
parameters from SDSS. We compared these types with those extracted 
from the HyperLeda database and found the mean scatter between 
these two systems of morphological classification being 2 $T$-units.

\begin{figure}
\centering
\includegraphics[width=6.0cm, angle=0, clip=]{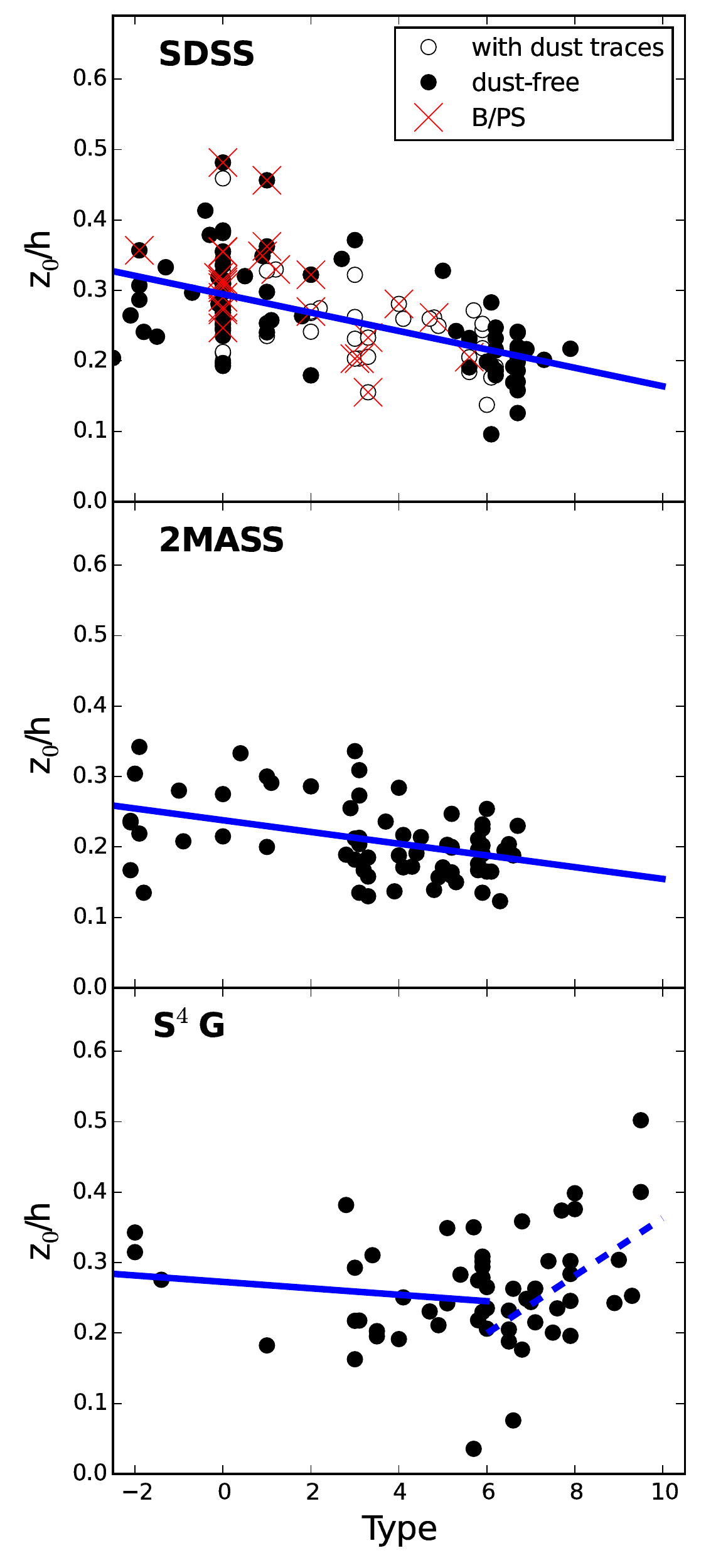}
\caption{The distribution of the disc flatness as a function of 
 the morphological type for the SDSS (top panel, $i$ band), 
 2MASS (middle panel, $K_\mathrm{s}$ band) and S$^4$G (bottom panel, 3.6~$\mu$m) sample. 
 The open circles in the top panel refer to the subsample of galaxies 
 with slightly visible dust lanes (59 galaxies). 
 The filled circles refer to the reference subsample of galaxies 
 without dust lanes (82 galaxies). The red crosses mark the 
 galaxies with sharp X-pattern structures in the central region. The blue lines represent the regression lines for each sample (see text).}
\label{z0h_Type}
\end{figure}

Fig.~\ref{z0h_Type} shows the distribution of the galaxies from our
SDSS, 2MASS and S$^4$G samples in the same $z_0/h$--$Type$ space. 
The SDSS sample is shown in Fig.~\ref{z0h_Type}, the top panel. 
We also marked here the galaxies with dust traces (open circles) and 
with visible X-shaped structure (red crossed symbols) which are 
candidates to hosting bars as discussed by \cite{Bureau+2006}.
% The greater part of the galaxies with the X-shaped structure have slightly visible dust lanes, 
% which can be explained by a specific bar-driven model with significant 
% dust production.
The whole sample demonstrates a trend (the Pearson's 
correlation coefficient $\rho=-0.56$, the two-sided p-value $p=1.44$e-12). The difference in 
the thickness between the thickest and the thinnest discs is of the 
factor of 1.5.

For the 2MASS sample of large galaxies we do not see strong 
variations of the relative thickness with the morphological type 
($\rho=-0.41$, $p=0.001$; see Fig.~\ref{z0h_Type}, middle plot). 
Since the S$^4$G sample consists mainly of late-type edge-on galaxies 
(many of them occur to be irregular), we have even an upward trend 
for the galaxies with $T>6$ (blue dashed regression line with $\rho=0.48$ 
and $p=0.016$ on Fig.~\ref{z0h_Type}, bottom panel). For the galaxies 
with $T\leq6$, the correlation between $z_0/h$ and $Type$ is not seen 
($\rho=-0.15$, $p=0.443$). Notwithstanding small number of early-type 
galaxies and excluding irregular galaxies with $T\geq8$, the overall 
picture does not reveal any correlation between the morphological 
type and the stellar disc flattening.
%Thus, we can not confirm the existence of a prominent 
% trend for this sample as well.

The poor dependence between the disc flattening and the galactic 
morphological type is confirmed indirectly by \cite{Chudakova+2014}. 
The latter paper presents a new photometric method for deriving the relative
thickness of the stellar disc from 2D surface brightness photometry. 
The method was applied to a sample of 45 early-type (S0-Sb) galaxies. 
They found that the discs of lenticular and early-type spiral 
galaxies have similar thickness, on average.

%%%%%%%%%%%%%%%%%%%%%%%%%%%%%%%%%%%%%%%%%%%%%%%%%%%%%%%%%%%%%%%%%%%%%%
%%%% Fig_9
%%%%%%%%%%%%%%%%%%%%%%%%%%%%%%%%%%%%%%%%%%%%%%%%%%%%%%%%%%%%%%%%%%%%%%
%%%%%%%%%%%%%%%%%%%%%%%%%%%%%%%%%%%%%%%%%%%%%%%%%%%%%%%%%%%%%%%%%%%%%%

%%%%%%%%%%%%%%%%%%%%%%%%%%%%%%%%%%%%%%%%%%%%%%%%%%%%%%%%%%%%%%%%%%%%%%
\subsection[]{The disc thickness versus the bulge-to-total luminosity ratio}
\label{res2}
%%%%%%%%%%%%%%%%%%%%%%%%%%%%%%%%%%%%%%%%%%%%%%%%%%%%%%%%%%%%%%%%%%%%%%
Determination of the morphological type for edge-on galaxies 
is difficult.  \cite{Bergh1998} in his review on the 
galactic morphology and classification stressed that edge-on galaxies 
are often difficult or impossible to classify. \cite{Naim+1995} 
concluded that the agreement between independent observers to do 
morphological classification is good, on average, but the scatter is 
large (about 1.8 subtypes). However, in the case of edge-on galaxies the 
agreement between independent classifiers is worse 
(see fig.~6, right panel in that work). In Section~\ref{disc2} 
we will demonstrate how the inclination of a galaxy can affect 
its morphological classification.

For edge-on galaxies it is more reliable to use the bulge-to-total 
luminosity ratio $B/T$ as the indicator of the galaxy morphological 
type because this value can be estimated directly from the 
photometric analysis and modeling.

\begin{figure}
\centering
\includegraphics[width=6.0cm, angle=0, clip=]{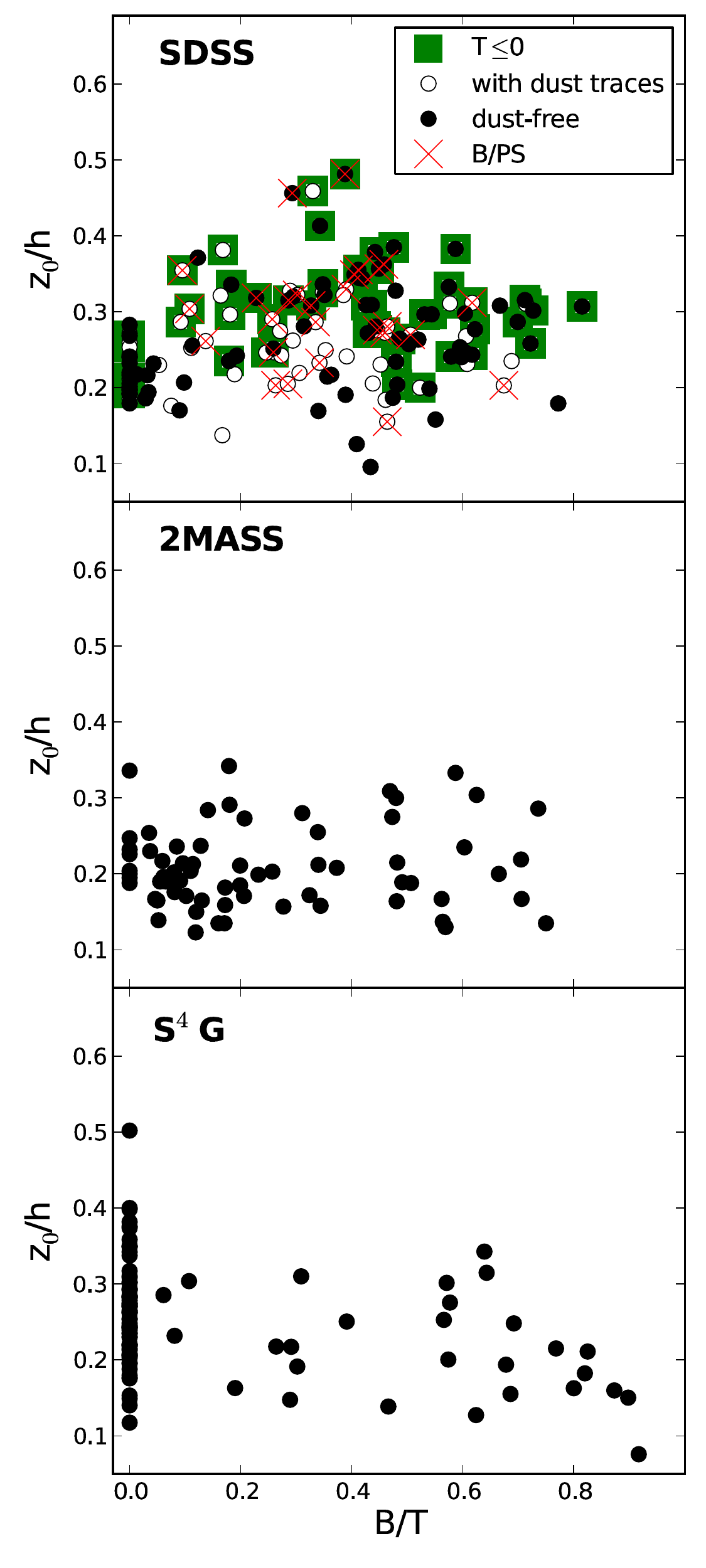}
\caption{The distribution of the stellar disc flatness as a 
 function of the bulge-to-total luminosity ratio for the SDSS 
 ($i$ band), 2MASS ($K_\mathrm{s}$ band) and S$^4$G (3.6~$\mu$m) samples. 
 Symbols as in Fig.~\ref{z0h_Type}, and the green squares represent early-type galaxies with $Type\leq0$.}
% The open circles refer to the 
% subsample of galaxies with slightly visible dust lanes (59 galaxies), 
% the filled circles refer to the reference subsample of galaxies 
% without dust lanes (82 galaxies). The red crosses mark the 
% galaxies with sharp X-pattern structure in the central region.}
\label{z0h_BT}
\end{figure}

In Fig.~\ref{z0h_BT} we plotted the dependence of the disc relative 
thickness $z_0/h$ on the bulge-to-total luminosity ratio $B/T$. 
One can see that for galaxies from SDSS (the top panel) with 
$0.2\lesssim B/T \lesssim0.6$ there is a large scatter of $z_0/h$ 
values around $z_0/h\approx0.3$. The galaxies 
with visible X-shaped structures (red crosses) 
lie within the same limits $0.2\lesssim B/T \lesssim0.6$ with the 
relative thickness $z_0/h\approx0.30\pm0.05$ in comparison with the average flatness for the whole sample $\langle z_0/h \rangle =0.25\pm0.07$. Thus, we may conclude that the galaxies with X-shaped structures have thicker stellar discs, on average. 
% Moreover, they all 
% have slightly visible dust lanes (open circles in Fig.~\ref{z0h_BT}). 
% Dust-free galaxies (black circles) have a bit smaller scatter in 
% the relative thickness but also do not show any trend with the 
% ratio $B/T$. 

We verified whether our fitting procedure could affect the 
possibly existed correlation between $z_0/h$ and $B/T$ taking into 
account the fact that these retrieved parameters may have 
10\% and 25\% uncertainties respectively 
(see Sect.~\ref{Simulations}). We created a simulated sample with 
the same distribution of $z_0/h$ as for the SDSS sample, where 
there exists correlation between the disc flatness and the 
bulge-to-total luminosity ratio with the regression coefficient 
$\rho=0.55$. We then applied the biasing to each point in the 
$B/T$--$z_0/h$ space using the normal distributions with the 
standard deviations $0.25\cdot B/T$ and $0.1\cdot z_0/h$ for the values 
of $B/T$ and $z_0/h$ respectively. We repeated this biasing 100000 
times and found that the probability to receive the final sample 
with the uncorrelated ($\rho<0.15$) parameters $z_0/h$ and $B/T$ is 
$5\cdot10^{-5}$. 
This test proves that the lack of the correlation between the disc 
flatness and the bulge-to-total luminosity ratio can not be caused 
by the fitting biases.

We also marked by squares early-type galaxies with $Type\leq0$
(Fig.~\ref{z0h_BT}, top plot). 
All these galaxies show a wide range of the ratio $B/T$ from 0 to 0.8. 
They also demonstrate a significant scatter of $z_0/h$ whereas the ratio 
$z_0/h$ does not depend on $B/T$ for these objects in accordance 
with the conclusion by \cite{Chudakova+2014}. The dependence 
in Fig.~\ref{z0h_Type} (top plot) is mainly created by these galaxies. 
The classification of these galaxies as S0 is expected to be biased 
because we can not conclude the absence of a spiral structure in them. The number of galaxies classified as early-type galaxies 
in 2MASS and S$^4$G samples, in turn, is small. As a result, the dependence of 
$z_0/h$ on $Type$ for these samples is weak (Fig.~\ref{z0h_Type}, middle and bottom 
plots).

For both 2MASS and S$^4$G samples we did not find any correlation between 
the flattening $z_0/h$ and the ratio $B/T$ as well. 
%We can see that in the samples very thin galaxies 
%of early as well as of late types are present. 
From the shown distribution we can conclude again 
that the early-type galaxies may have stellar disc with different 
flatness, including very thin discs.  
The correlation between the relative thickness 
and the bulge contribution to the total luminosity is not confirmed 
for our samples.

%%%%%%%%%%%%%%%%%%%%%%%%%%%%%%%%%%%%%%%%%%%%%%%%%%%%%%%%%%%%%%%%%%%%%%
%%%% Fig_
%%%%%%%%%%%%%%%%%%%%%%%%%%%%%%%%%%%%%%%%%%%%%%%%%%%%%%%%%%%%%%%%%%%%%%
% \begin{figure}
% \centering
% \includegraphics[height=3.0in, angle=0, clip=]{z0h_BT_2D.pdf}
% \caption{The distribution of the stellar disc flatness as a function 
% of the bulge-to-total luminosity plotted for the SDSS sample, in the 
% $i$ band. The open circles refer to the 
% subsample of galaxies slightly visible dust lanes (99 galaxies), 
% the filled circles refer to the reference subsample of galaxies 
% without dust lanes (46 galaxies). The red crosses mark the 
% galaxies with sharp X-pattern structure in the central region.}
% \label{z0h_BT_2D}
% \end{figure}
%%%%%%%%%%%%%%%%%%%%%%%%%%%%%%%%%%%%%%%%%%%%%%%%%%%%%%%%%%%%%%%%%%%%%%

%%%%%%%%%%%%%%%%%%%%%%%%%%%%%%%%%%%%%%%%%%%%%%%%%%%%%%%%%%%%%%%%%%%%%%
%%%% Fig_10
%%%%%%%%%%%%%%%%%%%%%%%%%%%%%%%%%%%%%%%%%%%%%%%%%%%%%%%%%%%%%%%%%%%%%%
%%%%%%%%%%%%%%%%%%%%%%%%%%%%%%%%%%%%%%%%%%%%%%%%%%%%%%%%%%%%%%%%%%%%%%

%%%%%%%%%%%%%%%%%%%%%%%%%%%%%%%%%%%%%%%%%%%%%%%%%%%%%%%%%%%%%%%%%%%%%%
\subsection[]{The disc thickness versus the galaxy colour}
\label{res3}
%%%%%%%%%%%%%%%%%%%%%%%%%%%%%%%%%%%%%%%%%%%%%%%%%%%%%%%%%%%%%%%%%%%%%%
One of the prominent correlations along the Hubble sequence is the 
morphological dependence on the integrated colour (see the review by 
\citealp{Roberts+1994}). The early-type galaxies are mostly red 
whereas the late-type ones are mostly blue. Thus, the colour, 
in addition to the ratio $B/T$, is another indicator of the 
morphological type. The dispersion of the relation between the 
colour and the morphological type is large: there are some galaxies 
of Sc type which are red, and vice versa, there may be blue S0/a 
galaxies. As an example, among the sample of edge-on galaxies 
we found FGC~1597 (Scd) to be red and NGC~3600 (Sa) to be blue. 

\begin{figure*}
\centering
\includegraphics[width=18.0cm, angle=0, clip=]{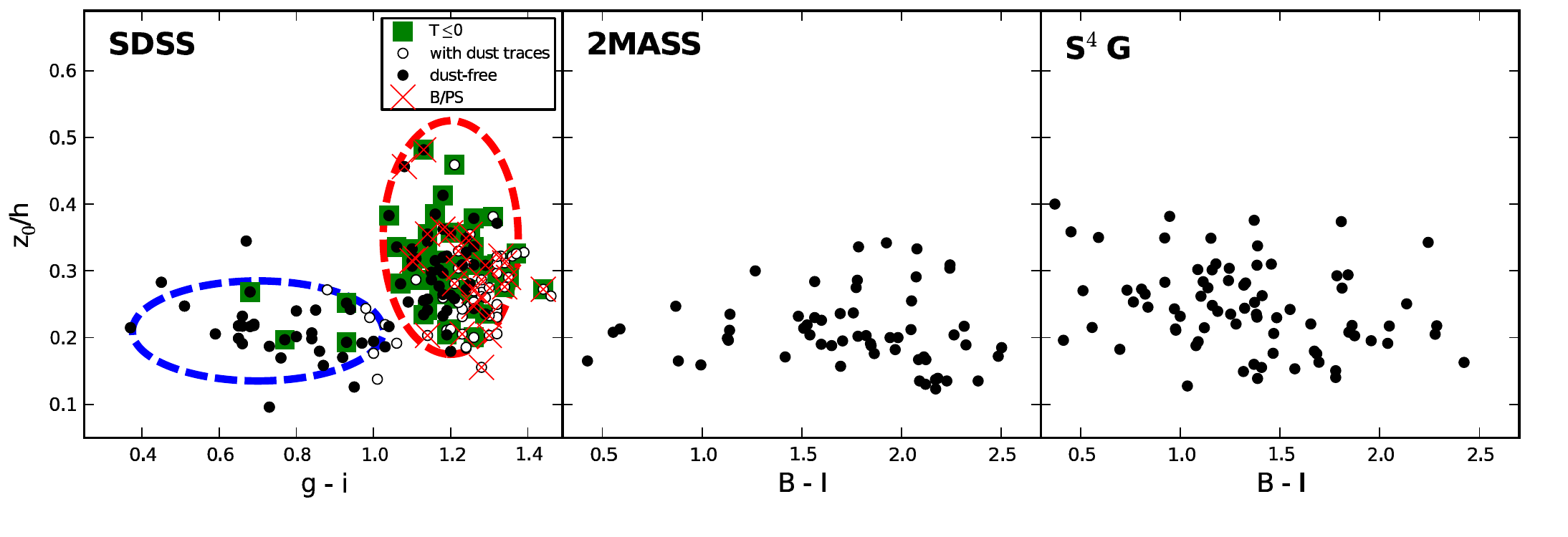}
\caption{The distribution of the disc flattening as a function of 
 the colour $g-i$ for the SDSS ($i$ band), 2MASS ($K_\mathrm{s}$ band) 
 and S$^{4}$G (3.6~$\mu$m) sample. The red and blue ovals outline the 
 two possible families of galaxies selected from their loci 
 on the $g-i$---$z_0/h$ plane. The symbol designation is kept the same 
 as in Fig.~\ref{z0h_BT}.} 
\label{z0h_col}
\end{figure*}

In fig.~7 from BKM14 there is no significant trend in the disc 
thickness with the overall galaxy colour $g-r$. 
In Fig.~\ref{z0h_col} we plot the $z_0/h$ versus 
the $g-i$ for the SDSS sample (left plot). 
The colours were corrected for the Milky Way reddening using the 
extinction maps by \cite{Schlegel+1998}, but they were not corrected 
for the internal extinction in the galaxies.

Fig.~\ref{z0h_col} (left plot) shows that there are two perpendicularly oriented
clouds of galaxies. The bluer galaxies ($g-i\lesssim1.0$) have stellar discs 
thinner ($z_0/h\approx0.21\pm0.03$) than the red galaxies, 
which additionally reveal a wide scatter of the thickness 
($z_0/h\approx0.28\pm0.06$). It is well seen that there are red galaxies with both flat and thickened discs. 

The latter can be interpreted in two ways. First, red galaxies 
(representing mostly early and intermediate types) can truly have 
thick or thin discs (as we could see in Fig.~\ref{galaxies_ex}). 
Second, the dust lane (and the internal extinction in general) can 
cause the reddening. 
We tried to properly mask the dust lane while preparing the image 
for the decomposition. But remaining traces of the dust lane can alter the profile 
in the $z$-direction making it shallower. It results in a larger value of the
$z_0$. In the latter case the results of the decomposition 
should be biased, and the relative thickness $z_0/h$ should be slightly overestimated. 
Almost all galaxies with the dust traces fall into the red 
cloud in the plot, while, as we can see, the dust-free galaxies fall 
into both clouds (Fig.~\ref{z0h_col}, filled circles).
% The red cloud 
% with the dust-free galaxies is not so prominent as the red cloud with 
% the galaxies demonstrating dust traces.
Note that we can not exclude biased decomposition in the $i$ band even in the case 
of dust-free galaxies since we do not see the division into two clouds 
when using SDSS sample for the ratio $B/T$ and $Type$. The galaxies with 
dust lanes sample the whole range of value $B/T$ and $Type$. 
It indicates that the presence of the dust lanes marginally correlates with 
the $B/T$ and $Type$ (see, for example,~\citealp{Draine+2007}).

For both 2MASS and S$^4$G samples there is no correlation between 
the disc thickness and the overall galaxy colour $B-I$ (middle and 
right plots). We can see only a slightly larger scatter for the ratio 
$z_0 / h$ in the redder side of the plot for the 2MASS sample.

Thus, the difference between the distributions of the parameters
estimated from the optical SDSS and IR samples
on the $z_0/h$--$colour$ diagram may be 
interpreted by the presence of the dust component that affects 
the model decomposition. We suppose that even our `dust-free' 
galaxies are not absolutely transparent in the $i$ band, which  
biases the $z_0/h$ estimation. However, this issue 
should be considered more carefully.

% Also, we should stress here that all three samples are not statistically complete.
Another possible factor that biases interpretations of the plots is that the S$^4$G sample 
is deficient in early-type red galaxies and comprises a large fraction of blue galaxies. 
This fact can be one of the reasons why we do not see any subdivision of galaxies 
in the $z_0/h$--$B-I$ space for the S$^4$G sample.

%%%%%%%%%%%%%%%%%%%%%%%%%%%%%%%%%%%%%%%%%%%%%%%%%%%%%%%%%%%%%%%%%%%%%%
%%%% Fig_11
%%%%%%%%%%%%%%%%%%%%%%%%%%%%%%%%%%%%%%%%%%%%%%%%%%%%%%%%%%%%%%%%%%%%%%
%%%%%%%%%%%%%%%%%%%%%%%%%%%%%%%%%%%%%%%%%%%%%%%%%%%%%%%%%%%%%%%%%%%%%%

%%%%%%%%%%%%%%%%%%%%%%%%%%%%%%%%%%%%%%%%%%%%%%%%%%%%%%%%%%%%%%%%%%%%%%
\subsection[]{The disc thickness versus galactic luminosity}
\label{res4}
\begin{figure*}
\includegraphics[width=18.0cm, angle=0, clip=]{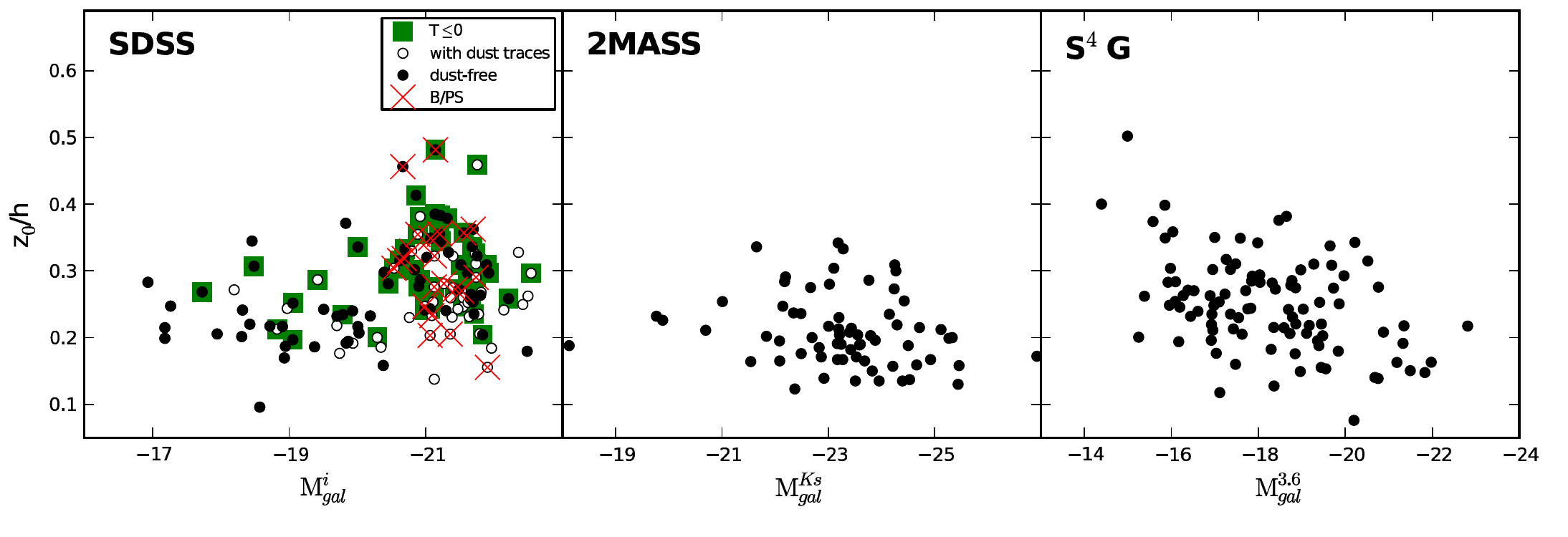}
\caption{The dependence of the disc relative thickness on the total 
 luminosity for the SDSS ($i$ band), 2MASS ($K_\mathrm{s}$ band) and S$^{4}$G (3.6~$\mu$m) 
 samples. The designation is kept the same as in Fig.~\ref{z0h_BT}.}
\label{z0h_Mgal}
\end{figure*}

%%%%%%%%%%%%%%%%%%%%%%%%%%%%%%%%%%%%%%%%%%%%%%%%%%%%%%%%%%%%%%%%%%%%%%
We also analyze the dependence of the relative thickness $z_0/h$ on 
the total luminosity of galaxies and find this correlation is 
weak if at all existing (see Fig.~\ref{z0h_Mgal}). 
Our SDSS sample demonstrates a large scatter for the ratio 
$z_0/h$ in the bright side of the plot (Fig.~\ref{z0h_Mgal}, left 
panel), which is similar to the presence of two perpendicular clouds 
in the $z_0/h$ --- $g-i$ diagram (see previous section). 
This scatter is mainly due to the presence of early-type 
galaxies ($Type \leq 0$, squares in the plot), which show a large 
scatter of the ratio $z_0/h$.

It is well established that the dust content correlates well with 
the galaxy mass (e.g.~\citealp{Masters+2010,Skibba+2011}), and we 
see that the bright region is populated mainly by the galaxies with 
dust traces. 

Only one sample (S$^4$G) shows a hint of the thickness -- luminosity 
relation: more luminous galaxies are thinner, on average. 
\cite{deGrijs1998} also did not find any evidence for a dependence 
between the relative thickness and the absolute 
magnitude (see fig.~7b in \citealp{deGrijs1998}). He referred 
to the theoretical prediction that the ratio $h/z_0$ 
should decrease rapidly from faint galaxies to a constant level for
normal and bright galaxies \citep{Bottema1993} and claimed that 
it could not be confirmed observationally in his sample. Our samples 
also do not confirm this trend. Moreover, S$^4$G sample shows the 
inverse trend probably because of the large number of late-type 
(irregular) galaxies. 
However this trend is consistent with the issue of the 
thickening of disc towards low mass reported in recent years in 
the literature \citep{Sanchez+2010,Roychowdhury+2013}.

%%%%%%%%%%%%%%%%%%%%%%%%%%%%%%%%%%%%%%%%%%%%%%%%%%%%%%%%%%%%%%%%%%%%%%
%%%% Fig_12
%%%%%%%%%%%%%%%%%%%%%%%%%%%%%%%%%%%%%%%%%%%%%%%%%%%%%%%%%%%%%%%%%%%%%%
%%%%%%%%%%%%%%%%%%%%%%%%%%%%%%%%%%%%%%%%%%%%%%%%%%%%%%%%%%%%%%%%%%%%%%

%%%%%%%%%%%%%%%%%%%%%%%%%%%%%%%%%%%%%%%%%%%%%%%%%%%%%%%%%%%%%%%%%%%%%%
\section{Discussion}
\label{Discussion}
%%%%%%%%%%%%%%%%%%%%%%%%%%%%%%%%%%%%%%%%%%%%%%%%%%%%%%%%%%%%%%%%%%%%%%

%%%%%%%%%%%%%%%%%%%%%%%%%%%%%%%%%%%%%%%%%%%%%%%%%%%%%%%%%%%%%%%%%%%%%%
\subsection[]{The difference between the 1D and 2D decomposition results}
\label{diff_dec}
\begin{figure}
\centering
\includegraphics[width=6.0cm, angle=0, clip=]{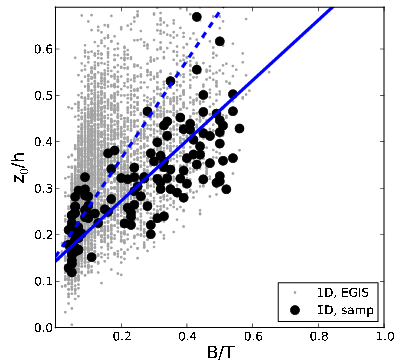}
\caption{The distribution of the disc flatness as a function of the 
 bulge-to-total luminosity ratio for the whole EGIS sample (gray dots) 
 and for the subsample of 141 galaxies (black circles) estimated using 
 the 1D decomposition in the $i$ band. 
 The dashed regression line is referred to the whole 
 sample; the solid regression line is referred to the 
 subsample.}
\label{z0h_BT_biz}
\end{figure}

%%%%%%%%%%%%%%%%%%%%%%%%%%%%%%%%%%%%%%%%%%%%%%%%%%%%%%%%%%%%%%%%%%%%%%
The dependence of the stellar disc relative thickness on the 
$Type$ reveals itself clearly only in the 1D modeling 
(\citealp{deGrijs1998}; BKM14). This statement is also applicable to  
the dependence between the disc flatness $z_0/h$ and the bulge-to-total luminosity 
ratio $B/T$. Fig.~\ref{z0h_BT_biz} demonstrates this dependence 
for the SDSS sample in the case of 1D analysis. The correlation 
between the disc thickness and $B/T$ is good (the Pearson's 
correlation coefficient $\rho=0.75$) for the subsample of 141 
galaxies (black circles), and it is fairly good 
for the whole EGIS catalogue (gray dots, $\rho=0.53$). 
However, the correlation does not appear at all 
if the 2D analysis results are used
(see Fig.~\ref{z0h_BT}, top panel).

%%%%%%%%%%%%%%%%%%%%%%%%%%%%%%%%%%%%%%%%%%%%%%%%%%%%%%%%%%%%%%%%%%%%%%
%%%% Fig_13
%%%%%%%%%%%%%%%%%%%%%%%%%%%%%%%%%%%%%%%%%%%%%%%%%%%%%%%%%%%%%%%%%%%%%%
%%%%%%%%%%%%%%%%%%%%%%%%%%%%%%%%%%%%%%%%%%%%%%%%%%%%%%%%%%%%%%%%%%%%%%

\begin{figure*}
\centering
\includegraphics[width=13.0cm, angle=0, clip=]{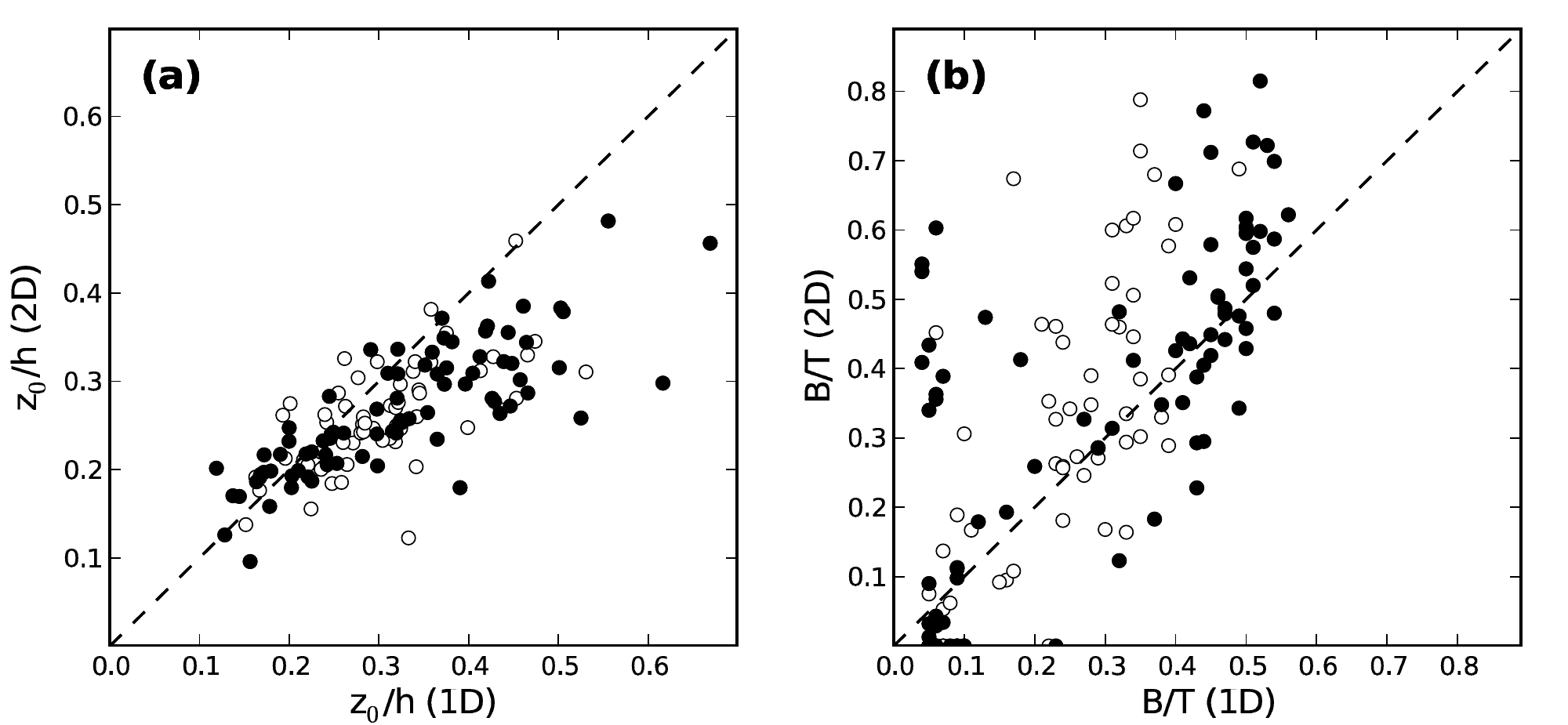}
\caption{The comparison of the results for the $z_0/h$ and $B/T$ 
 approximation between the 1D and 2D decomposition techniques. 
 The open circles refer to the subsample of galaxies with slightly 
 visible dust lanes; the filled circles refer to the 
 reference subsample of galaxies without dust lanes.} 
\label{compar_1D_2D}
\end{figure*}

The correlation between the $z_0/h$ and $B/T$ parameters for the 
1D analysis of the SDSS subsample galaxies can be explained by 
biased estimation of the disc structural parameters for the galaxies with 
significant bulge contribution. This is rather expected because in 
the early-type galaxies the extension of the bulge structure is much 
larger than that in late-type galaxies. The disc scale height $z_0$ 
should be overestimated while analyzing the vertical photometric disc 
profiles contaminated by the bulge light. At the same time, the radial 
scale length should be underestimated. As a result, the ratio 
$z_0/h$ will be overestimated in the early-type galaxies. The biased 
1D decomposition leads also to the underestimation of the ratio 
$B/T$ in comparison with 2D decomposition. 
In Fig.~\ref{compar_1D_2D} we compare the results for the 
$z_0/h$ and $B/T$ approximation between the 1D and 2D 
decomposition techniques applied to our SDSS sample. The ratio $z_0/h$ is overestimated for the 1D 
decomposition in comparison with the 2D decomposition (left panel). 

The biased decomposition in the 1D analysis reveals itself in the 
biased distribution of the ratio $B/T$ (Fig.~\ref{z0h_BT_biz} 
and Fig.~\ref{compar_1D_2D}, right panel) 
in the comparison with the distribution over $B/T$ in the 2D analysis 
($B/T < 0.6$ for the 1D decomposition, and $B/T$ spreads up to 0.8 for 
the 2D decomposition for the same galaxies). In the 1D analysis the 
contribution of the bulge seems to be underestimated and the rest 
of the true bulge light (non-retrieved from the profile) contributes to 
the disc thickness.

%%%%%%%%%%%%%%%%%%%%%%%%%%%%%%%%%%%%%%%%%%%%%%%%%%%%%%%%%%%%%%%%%%%%%%
%%%% Fig_14
%%%%%%%%%%%%%%%%%%%%%%%%%%%%%%%%%%%%%%%%%%%%%%%%%%%%%%%%%%%%%%%%%%%%%%
%%%%%%%%%%%%%%%%%%%%%%%%%%%%%%%%%%%%%%%%%%%%%%%%%%%%%%%%%%%%%%%%%%%%%%% 

\begin{figure}
\centering
\includegraphics[width=5.7cm, angle=0, clip=]{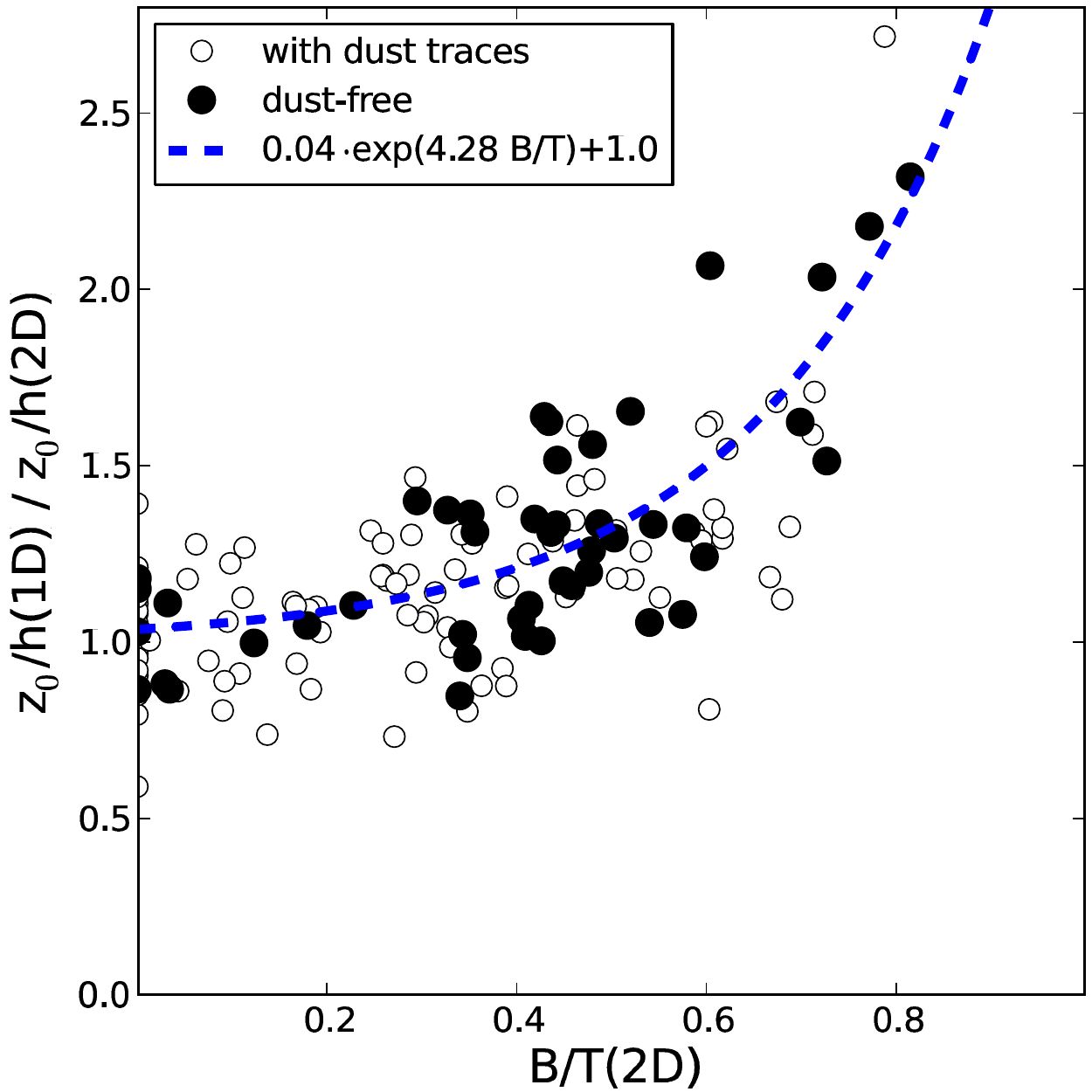}
\caption{The dependence of the ratio of the stellar thicknesses in 
 the 1D and 2D modeling versus the bulge-to-total luminosity ratio 
 from the 2D analysis for our SDSS sample, $i$ band. 
 The open circles refer to the subsample of galaxies with slightly 
 visible dust lanes; the filled circles refer to the 
 reference subsample of galaxies without dust lanes. The dashed line 
 is a simple exponential approximation to the data.} 
\label{z0h_rel_BT}
\end{figure}

In Fig.~\ref{z0h_rel_BT} the ratio of the stellar disc thickness
in the 1D and 2D analysis versus the bulge-to-total luminosity ratio 
(from the 2D analysis) is shown. It is well seen that with the 
increasing of the $B/T$ the relative thickness is growing with 
respect to the 2D approach (both for the whole sample and 
the reference subsample of 46 dust-free galaxies). In other 
words, the correlation in Fig.~\ref{z0h_BT_biz} may be a 
consequence of the biased model fitting in the 1D analysis, which 
dramatically affects the decomposition results for galaxies with 
significant non-disc components. 
A simple exponential approximation of data is shown in 
Fig.~\ref{z0h_rel_BT} by dashed line. We can see that for 
galaxies with $B/T\gtrsim0.4$ the 1D disc flattening will overestimate 
more than 10\% the flattening found with the the 2D method and more than twice for galaxies with $B/T\approx0.8$.

%%%%%%%%%%%%%%%%%%%%%%%%%%%%%%%%%%%%%%%%%%%%%%%%%%%%%%%%%%%%%%%%%%%%%%
%%%% Fig_15
%%%%%%%%%%%%%%%%%%%%%%%%%%%%%%%%%%%%%%%%%%%%%%%%%%%%%%%%%%%%%%%%%%%%%%
%%%%%%%%%%%%%%%%%%%%%%%%%%%%%%%%%%%%%%%%%%%%%%%%%%%%%%%%%%%%%%%%%%%%%%

We summarize that the 1D and 2D methods produce different 
results as we can see by comparing Fig.~\ref{z0h_BT_biz} and 
Fig.~\ref{z0h_BT} (top panel). 
We assume that this difference is caused 
by biased estimation of the bulge contribution to the total luminosity in the 1D 
approach. 
With the 1D analysis, the bulge was not fitted at all and only its fraction 
was estimated which may bring wrong decomposition results 
for bulge-dominated galaxies. The 2D approach is a more robust method, which should report reliable results.
The 3D analysis performed in BKM14 seems to be rather 
ambiguous, and it should be properly improved. Therefore, we do 
not discuss the results of this method here.

%%%%%%%%%%%%%%%%%%%%%%%%%%%%%%%%%%%%%%%%%%%%%%%%%%%%%%%%%%%%%%%%%%%%%%
\subsection[]{Morphological type estimation}
\label{disc2}
%%%%%%%%%%%%%%%%%%%%%%%%%%%%%%%%%%%%%%%%%%%%%%%%%%%%%%%%%%%%%%%%%%%%%%

The morphological classification of edge-on galaxies is a difficult task
because of several aspects. First, the most essential 
criterion of the Hubble type, the tightness of the spiral pattern, 
cannot be applied to galaxies viewed edge-on. Second, dust 
attenuation dims the bulge in optics and NIR and hence, we are not able to 
estimate by eye the contribution of the bulge to the total 
luminosity in the galaxy reliably. Third, some structural details in 
a galaxy can be identified only via the inspection of its non-edge-on 
surface brightness maps. For instance, steps (``shoulders'') 
at the photometric profile along the major axis of an edge-on galaxy 
can be an evidence of a bar or a ring, but they can also reveal the deviation of the 
disc surface brightness profile from the single exponent. Wrong 
morphological classification has been found for many galaxies 
when detailed bulge/disc decomposition was performed 
(see, for example, \citealp{desouza+2004}).

\begin{figure}
\centering
\includegraphics[width=5.7cm, angle=0, clip=]{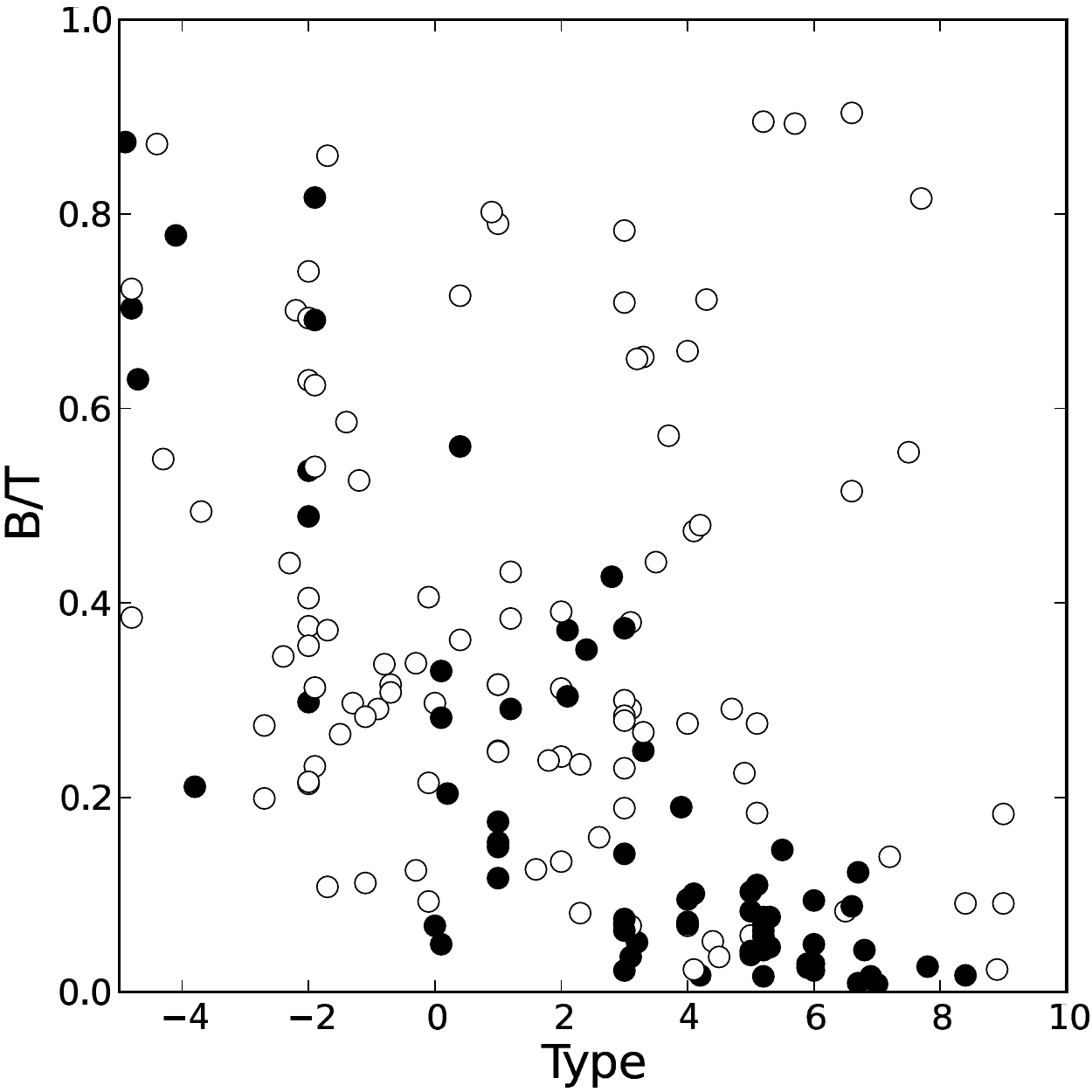}

\caption{The distribution of the bulge-to-total luminosity ratio as 
 a function of the morphological type for galaxies with inclination 
 angles $i=90^{\circ}$ (open circles) and $i<30^{\circ}$ 
 (black circles), in the S$^4$G sample (3.6~$\mu$m).}
\label{BT_Type_incl}
\end{figure}

In order to demonstrate how the inclination of a galaxy can affect 
its morphological classification, we show the dependence between 
the $B/T$ and $Type$ for two subsamples from the S$^4$G catalogue 
(Fig.~\ref{BT_Type_incl}). The first group contains galaxies with 
$i<30^{\circ}$ (black circles). The second one comprises galaxies 
with $i=90^{\circ}$ (open circles). Here, the inclination angle 
$i$ is taken from the HyperLeda database. Although some of these 
galaxies are not viewed perfectly edge-on, they are certainly 
highly inclined. 
In Fig.~\ref{BT_Type_incl} one can see that the open circles have 
larger scatter, with many outliers in the right upper corner. 
This indicates that the morphological classification of 
edge-on galaxies is rather ambiguous. Observable strong dust 
attenuation can dim the bulge and hence bias the morphology 
classification for such objects towards the later types. 
%That is why we chose a subsample of galaxies without dust lanes 
% in SDSS to study effects of the morphological type.

%%%%%%%%%%%%%%%%%%%%%%%%%%%%%%%%%%%%%%%%%%%%%%%%%%%%%%%%%%%%%%%%%%%%%%
%%%% Fig_16
%%%%%%%%%%%%%%%%%%%%%%%%%%%%%%%%%%%%%%%%%%%%%%%%%%%%%%%%%%%%%%%%%%%%%%
%%%%%%%%%%%%%%%%%%%%%%%%%%%%%%%%%%%%%%%%%%%%%%%%%%%%%%%%%%%%%%%%%%%%%%

%%%%%%%%%%%%%%%%%%%%%%%%%%%%%%%%%%%%%%%%%%%%%%%%%%%%%%%%%%%%%%%%%%%%%%
\subsection[]{The model dependence of the obtained results}
\label{model_dependence}
%%%%%%%%%%%%%%%%%%%%%%%%%%%%%%%%%%%%%%%%%%%%%%%%%%%%%%%%%%%%%%%%%%%%%%
We should now make some comments on decomposing multicomponent 
galaxies. It is obviously difficult to unambiguously identify the presence of a bar in 
nearly edge-on systems. However, about half of disc galaxies demonstrates 
boxy/peanut-shaped structures in the central region (\citet{Lutticke+2000,Laurikainen+2011}), 
which is tagged 
by some authors as projection of the vertical structure of bars 
\citep{Bureau+1999,Chung+2004}. 

It is apparent that for such galaxies the simple `bulge+disc' model 
will not provide appropriate fitting. In this case a `bulge+disc+bar' 
modeling should be used instead. The bulge-to-total luminosity ratio 
and the relative disc thickness are expected to decrease in the case 
of the bar contribution. More detailed images of nearby galaxies 
show some other hidden features. For instance, the photometric decomposition of the Sombrero 
galaxy NGC~4594 revealed a disc, a bulge, an outer spheroid, 
a stellar ring, and a stellar, inner ring or disc 
\citep{Gadotti+2012}. 
However, in the case of such multi-component structure the fitted 
model becomes ambiguous and suffers from the local $\chi^2$ minimum 
degeneracy.

Also, some authors tried to find the parameters of thin and 
thick discs using a two-component model of the stellar disc 
(see, e.g. \citealp{Yoachim+2006,Comeron+2011}). In this work we only 
studied one-component disc decompositions (mainly because of the shallow 
photometric depth of the surveys we work with and a
lack of sufficient spatial resolution) following the works 
by \cite{deGrijs1998} and \cite{Kregel+2002}. Studying larger 
samples in different photometrical bands, we did not find a 
correlation between the disc flattening and the morphological 
type, as well as between the disc flattening and the bulge 
contribution to the total galaxy luminosity. 
%%Perhaps, these 
%%correlations will appear if thin and thick discs are included 
%%during decomposition of galaxy images of high resolution.

Our next step in studying edge-on galaxies is building the best 
models of nearby, well-resolved edge-on galaxies using their 
multi-band imaging, with proper dust-component treatment 
(see e.g. \citealt{Baes+2011}). 

\section{Conclusions}
\label{Conclusions}
%%%%%%%%%%%%%%%%%%%%%%%%%%%%%%%%%%%%%%%%%%%%%%%%%%%%%%%%%%%%%%%%%%%%%%

The results of our analysis of edge-on galaxies can be summarized 
as follows:
\begin{enumerate}

\item In the course of our analysis we obtained a 
counter-intuitive result, --- the disc relative thickness does not 
depend on the bulge-to-total luminosity ratio for all the samples 
studied. 
However, it may depend on the total luminosity in the near- 
and mid-infrared bands, in which the old stellar population 
contributes mainly in the disc luminosity, and reflects the dependence on 
the total mass \citep{Sanchez+2010,Roychowdhury+2013}. 
The existence of the correlation between the disc flatness and 
the morphological type is only confirmed for the SDSS sample 
and late-type galaxies from the S$^4$G sample (anti-correlation).

These findings may have far-reaching consequences.
Thus, since the relative thickness of the disc depends on 
the contribution of the spherical component (`bulge+dark halo') 
($h/z_0 \propto M_\mathrm{dark}/M_\mathrm{total}$; 
\citealp{Zasov+1991,Zasov+2002}, see also the discussion in 
MSR10 and \citealp{Mosenkov+2014}), 
this means that the relative contribution of 
the spherical component does not vary systematically along the 
morphological Hubble sequence of bright spiral galaxies. 
In turn, since bulges contribute more to the mass of the 
spherical component in the early-type disc galaxies, 
the dark halos themselves should reveal higher mass fraction in
the case of the late-type (mostly bulgeless) galaxies.

%%This conclusion is preliminary, of course, and it should be tested on 
%%independent observational data. 
At the same time, a compact but massive bulge stops the disc heating 
in the vertical direction due to the bending 
instability~\citep{Sotnikova2006} and leaves the disc rather thin. 
In galaxies with large $B/T$ ratio we may expect observing stellar 
discs of either thickness. In Fig.~\ref{z0h_BT} we see a large 
scatter of the ratio $z_0/h$ for the galaxies with $B/T > 0.4$.

\item We conclude that the correlation between the morphological 
type and the disc flatness reported earlier by some authors 
\citep[e.g.][]{deGrijs1998} was due to biased estimates of the 
disc flatness during the 1D analysis. Reanalyzing the same data by 
the 2D method \cite{Kregel+2002} found the disc scale height to 
be overestimated for almost all galaxies studied with the 1D 
decomposition (see their fig.~2, bottom panel). 
At the same time, the disc scale length was underestimated 
(the same fig.~2, top panel). As a result the relative thickness 
increases in the the 1D analysis. 
We obtained the same result and extended it: the 1D decomposition 
overestimates the disc thickness mainly in the galaxies with 
large bulges (more than 10\% for galaxies with $B/T\gtrsim0.4$), i.e. in all early-type disc galaxies. It is important 
to note that \cite{Kregel+2002} did not present the correlation 
between the disc thickness and the galaxy type for the updated 
analysis. It was done by \cite{Hernandez+2006} and is shown in their 
fig.~7. A glance at the plot suggests that the reported trend is poor 
and maintained by only one point representing a Sa type galaxy. 
Moreover, there are no galaxies earlier than the Sa in the sample by 
\cite{Kregel+2002}.

\item We also demonstrate that there is no correlation between the 
disc relative thickness and the galaxy colour. This fact strengthens 
our conclusion that there is no correlation between the thickness 
of the disc and the morphological type of the galaxy.

% \item There is no dependence between the disc relative thickness 
% and the bulge-to-total luminosity. This conclusion is confirmed 
% for all the samples in different passbands.
\end{enumerate}

%%%%%%%%%%%%%%%%%%%%%%%%%%%%%%%%%%%%%%%%%%%%%%%%%%%%%%%%%%%%%%%%%%%%%%
\section*{Acknowledgments}
%%%%%%%%%%%%%%%%%%%%%%%%%%%%%%%%%%%%%%%%%%%%%%%%%%%%%%%%%%%%%%%%%%%%%%
The authors express gratitude for the grant of the Russian 
Foundation for Basic Researches number 14-02-00810 and 14-22-03006-ofi. 
This work was partly supported by St. Petersburg State University 
research grant 6.38.669.2013. AM is a beneficiary of a mobility grant from the Belgian Federal Science Policy Office.
DB was partly supported by RSF via grant RSCF-14-22-00041.
%We are grateful to the referee for the comments and suggestions that 
%lead to significant improvements of this paper. 

We are thankful to the referee for his constructive
comments which helped improve the quality and the presentation of 
the paper.

Funding for the SDSS has been provided by the Alfred P. Sloan Foundation, the Participating Institutions, the
National Science Foundation, the U.S. Department of Energy, the National Aeronautics and Space Administration,
the Japanese Monbukagakusho, the Max Planck Society, and the Higher Education Funding Council for England. The
SDSS Web Site is http://www.sdss.org/.

This research makes use of the NASA/IPAC Extragalactic Database 
(NED) which is operated by the Jet Propulsion Laboratory, California 
Institute of Technology, under contract with the National 
Aeronautics and Space Administration, and the LEDA 
database (http://leda.univ-lyon1.fr). We also acknowledge the data products from the S$^{4}$G survey (http://www.cv.nrao.edu/$\sim$ksheth/S4G/Site/S4G\_Home.html) and the Two
Micron All Sky Survey, which is a joint project of the University of Massachusetts and the Infrared Processing and
Analysis Center/California Institute of Technology, funded
by the National Aeronautics and Space Administration and
the National Science Foundation.

\bibliographystyle{mn2e}
\bibliography{art}

%%%%%%%%%%%%%%%%%%%%%%%%%%%%%%%%%%%%%%%%%%%%%%%%%%%%%%%%%%%%%%%%%%%%%%
\appendix
\section{The estimation of the galaxy inclination}
\label{Appendix}
%%%%%%%%%%%%%%%%%%%%%%%%%%%%%%%%%%%%%%%%%%%%%%%%%%%%%%%%%%%%%%%%%%%%%%
In this work we use two different approaches to estimate nearly edge-on galaxy 
disc inclination which are presented in detail by 
Mosenkov (2015, in prep.).
%%%%%%%%%%%%%%%%%%%%%%%%%%%%%%%%%%%%%%%%%%%%%%%%%%%%%%%%%%%%%%%%%%%%%%
\begin{enumerate}
\renewcommand{\theenumi}{(\arabic{enumi})}
\item
The first approach is by using the $ExponentialDisk3D$ function in the {\sevensize IMFIT} 
code \citep{Erwin2014}. The code has an advantage over 
{\sevensize GALFIT} as it includes several 3D functions and the  
line-of-sight integration through a 3D luminosity-density model 
to create a projected 2D image. These functions are numerically 
integrated in each image pixel. The decomposition via the 3D integration 
is much slower than the 2D edge-on disc 
surface brightness evaluation~\eqref{form_DSB}. To speed up computations, 
we mask the central, bulge-dominated part in galaxies 
and use the $ExponentialDisk3D$ function in the {\sevensize IMFIT} to fit the periphery of the galaxies 
where the disc dominates. We chose the radius of the masking circle to be 
a half of the radius of the galaxy, which appears to be sufficient even for early-type galaxies. %We are convinced that such masking 
% works even for galaxies with high $B/T$. 

The model parameters in the $ExponentialDisk3D$ function include the disc exponential 
scale length $h$, the vertical scale height $z_0$, the central 
luminosity density $I(0,0)=I_{0}^{edge-on}/2h$ and the inclination 
angle of the disc $i$. In order to find the initial estimates 
of the galactic disc parameters, we use the {\sevensize DECA} code 
where the disc is described by (\ref{form_DSB}). These 
estimates are assumed as the initial values for running 
{\sevensize IMFIT}, in combination with the assumption of $i=90\degr$. The output of 
this code is the set of the disc structural parameters, including the 
estimated inclination angle $i$. 

In order to verify how well this method can estimate the 
inclination angle $i$ of the stellar disc, we performed the same 
simulations as in Section~\ref{Simulations}, in which instead of 
using~\eqref{form_DSB} for describing the edge-on disc, we used 
the $ExponentialDisk3D$ function with three different values of $i$: $80\degr$, $85\degr$ and 
$90\degr$. Then we fitted these simulated images with 
the $ExponentialDisk3D$ function of {\sevensize IMFIT}, where 
the initial disc parameters were estimated with the 
{\sevensize DECA} code, and the central region with the bulge 
was masked. The results of the inclination angle estimation 
using {\sevensize IMFIT} (the median value $\langle i \rangle$ 
and the standard deviation $\Delta i$) are compared with a priori known 
angles $i$:
\[
i=80\degr: \langle i \rangle=81.\degr6,\, \Delta i=3.\degr6\,,
\]
\[
i=85\degr: \langle i \rangle=85.\degr4,\, \Delta i=2.\degr5\,,
\]
\[
i=90\degr: \langle i \rangle=88.\degr9,\, \Delta i=1.\degr8\,.
\]

The above results prove that this method can be used to estimate 
the inclination angle in the disc galaxies without dust attenuation and/or 
in the near- and mid-infrared bands. We found that masking of the 
bulge essentially reduces the computational time while has no strong effect 
on the results of the disc decomposition. 

\item
The second method can be applied to the galaxies with visible dust lanes or 
with slight dust traces (similar to \citet{BizyaevKajsin2004}). 
%%Suppose, we detect a dust lane in the image of 
%%the nearly edge-on galaxy, and the plane of the dust distribution 
%%fully coincides with the stellar disc plane. 
We make a natural assumption that the mid-planes of the stellar and dust 
discs coincide in the galaxies. 
%%If we make a 
%%photometrical cut perpendicular to the disc plane and passing 
%%through the galaxy center, we can analyze the received 
%%photometrical profile and fit it with a function, for example, 
We analyze 1D vertical photometric profiles cut along the minor axis and fit them 
with the \ser\ function. The deficiency of light found by comparing 
the model and the observing profiles, and the asymmetry of the 
profile relative to the center of the galaxy indicate where the 
dust lane locates (its shift $\Delta z_\mathrm{dust}$ with respect
to the galactic mid-plane). 
Knowing the diameter of the galaxy $D$ and $\Delta z_\mathrm{dust}$, we can estimate the inclination angle 
of the galaxy $i$ (see Fig.~\ref{eon_incl}) as:
%%%%%%%%%%%%%%%%%%%%%%%%%%%%%%%%%%%%%%%%%%%%%%%%%%%%%%%%%%%%%%%%%%%%%%
\begin{equation}
i = 
  \arccos\left(\frac{2\,\Delta z_\mathrm{dust}}{D}\right)\,.
\label{incl}
\end{equation}
%%%%%%%%%%%%%%%%%%%%%%%%%%%%%%%%%%%%%%%%%%%%%%%%%%%%%%%%%%%%%%%%%%%%%%

%%%%%%%%%%%%%%%%%%%%%%%%%%%%%%%%%%%%%%%%%%%%%%%%%%%%%%%%%%%%%%%%%%%%%%
%%%% Fig_17
%%%%%%%%%%%%%%%%%%%%%%%%%%%%%%%%%%%%%%%%%%%%%%%%%%%%%%%%%%%%%%%%%%%%%%
\begin{figure}
\centering
\includegraphics[width=8cm, angle=0, clip=]{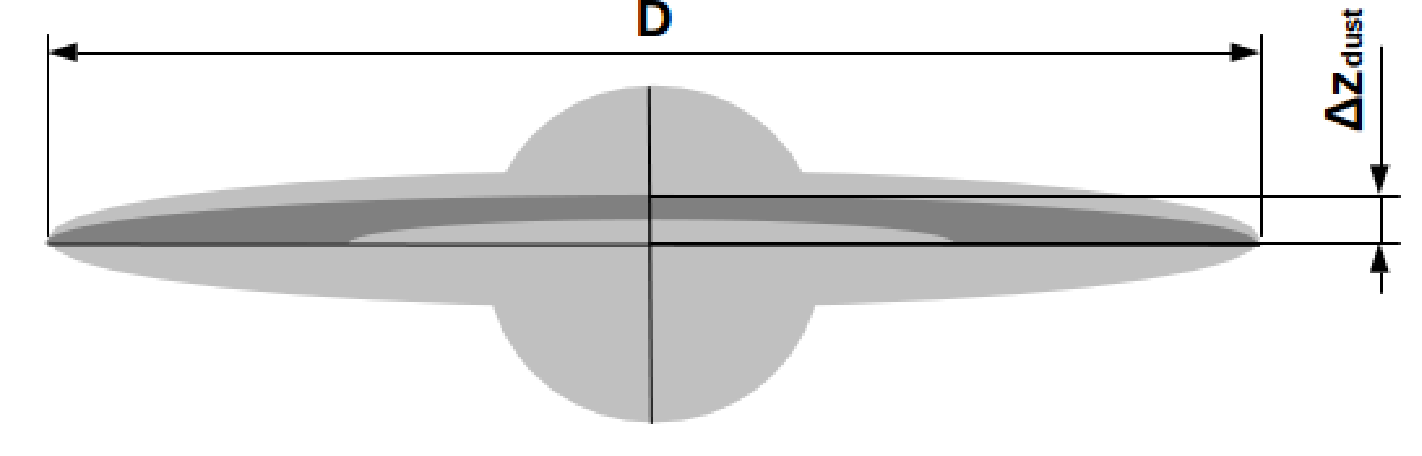}

\caption{A sketch of a nearly edge-on galaxy with the dust lane (see the text).}
\label{eon_incl}
\end{figure}
%%%%%%%%%%%%%%%%%%%%%%%%%%%%%%%%%%%%%%%%%%%%%%%%%%%%%%%%%%%%%%%%%%%%%%

%%%%%%%%%%%%%%%%%%%%%%%%%%%%%%%%%%%%%%%%%%%%%%%%%%%%%%%%%%%%%%%%%%%%%%
% \begin{equation}
% \Delta i = 
%   \frac{2\,\delta_\mathrm{dust}}
%   {D\,\sqrt{1-(2\,\Delta z_\mathrm{dust}/D)^2}}\,.
% \label{err_incl}
% \end{equation}
%%%%%%%%%%%%%%%%%%%%%%%%%%%%%%%%%%%%%%%%%%%%%%%%%%%%%%%%%%%%%%%%%%%%%%
\end{enumerate}
%%%%%%%%%%%%%%%%%%%%%%%%%%%%%%%%%%%%%%%%%%%%%%%%%%%%%%%%%%%%%%%%%%%%%%

\label{lastpage}

\end{document}